\documentclass[
superscriptaddress,
 amsmath,amssymb,
 pra,
twocolumn,
floatfix,
longbibliography
]{revtex4-1}

\usepackage{graphicx}
\usepackage{dcolumn}
\usepackage{bm}
\usepackage{comment}
\usepackage{mathrsfs}
\usepackage{subfig}
\usepackage[dvipdfmx]{hyperref}
\usepackage[usenames]{color}
\hypersetup{
 colorlinks=true,
 linkcolor=blue,
 citecolor=blue,
 urlcolor=blue
}

\newcommand{\Eqref}[1]{Eq.\,(\ref{#1})}
\newcommand{\re}{{\rm Re}\hspace{.2em}}
\newcommand{\sign}{{\rm sign}}

\begin{document}

\preprint{APS/123-QED}

\title{
Spin-1 Quantum Walks
}

\author{Daichi \surname{Morita}}
\affiliation{Graduate School of Pure and Applied Sciences, University of Tsukuba, 1-1-1 Tennodai, Tsukuba, Ibaraki 305-8571, Japan}

\author{Toshihiro \surname{Kubo}}
\affiliation{Graduate School of Pure and Applied Sciences, University of Tsukuba, 1-1-1 Tennodai, Tsukuba, Ibaraki 305-8571, Japan}

\author{Yasuhiro \surname{Tokura}}\email{tokura.yasuhiro.ft@u.tsukuba.ac.jp}
\affiliation{Graduate School of Pure and Applied Sciences, University of Tsukuba, 1-1-1 Tennodai, Tsukuba, Ibaraki 305-8571, Japan}
\affiliation{NTT Basic Research Laboratories, NTT Corporation, 3-1 Morinosato Wakamiya, Atsugi, Kanagawa 243-0198, Japan}

\author{Makoto \surname{Yamashita}}
\affiliation{Graduate School of Pure and Applied Sciences, University of Tsukuba, 1-1-1 Tennodai, Tsukuba, Ibaraki 305-8571, Japan}
\affiliation{NTT Basic Research Laboratories, NTT Corporation, 3-1 Morinosato Wakamiya, Atsugi, Kanagawa 243-0198, Japan}

\date{\today}

\begin{abstract}
We study the quantum walks of two interacting spin-1 bosons. We derive an exact solution for the time-dependent wave function that describes the two-particle dynamics governed by 
the one-dimensional spin-1 Bose-Hubbard model. We show that propagation dynamics in real space and mixing dynamics in spin space are correlated via the spin-dependent interaction in this system.  
The spin-mixing dynamics has two characteristic frequencies in the limit of large spin-dependent interactions. One of the characteristic frequencies is determined by the energy difference 
between two bound states, and the other frequency relates to the cotunneling process of a pair of spin-1 bosons. Furthermore, we numerically analyze the growth of the spin correlations in 
quantum walks. We find that long-range spin correlations emerge showing a clear dependence on the sign of the spin-dependent interaction and the initial state. 

\end{abstract}

\maketitle


\section{\label{sec:intro}Introduction}
Classical random walks play an important role in randomized algorithms that have been developed to achieve superior performance when solving various hard problems in computer science \cite{opac-b1089638}. 
It is thus quite natural that quantum walks (QWs) \cite{AMBAINIS:2003aa,doi:10.1080/00107151031000110776,Kendon:2007:DQW:1348911.1348916,Venegas-Andraca:2012:QWC:2386737.2386759}, 
which are the quantum mechanical counterparts of classical random walks, become a powerful tool for building quantum algorithms, providing versatile applications such as quantum search algorithms \cite{PhysRevA.67.052307,PhysRevA.70.022314} and universal quantum computation \cite{PhysRevLett.102.180501,PhysRevA.81.042330,Childs15022013}. Two theoretical QW models have already been proposed: the discrete-time QW \cite{Aharonov:2001:QWG:380752.380758,Ambainis:2001:OQW:380752.380757} and the continuous-time QW \cite{PhysRevA.58.915}. In discrete-time QWs, the dynamics of a {\it walker} is determined by flipping 
the coin state via a unitary operator at each discrete step. On the other hand, in continuous-time QWs, a {\it walker} evolves continuously on the basis of the Schr\"{o}dinger equation without flipping any coin states. These two models have revealed the unique features of QWs. A {\it walker} generates a coherent superposition state as a result of multiple interferences and propagates ballistically showing a bimodal profile of the probability distribution, which is in sharp contrast to classical random walks.  

Implementations of QWs have been reported in a series of experiments using magnetic resonance, trapped ions, trapped neutral atoms, and some photonic systems \cite{Manouchehri:2013:PIQ:2566741}. 
In particular, in recent years, continuous-time QWs including two {\it walkers} (i.e., two indistiguishable particles) have been attracting considerable attention \cite{PhysRevLett.102.253904,Peruzzo1500,PhysRevA.86.011603,Preiss13032015}.  
Experiments with an array of coupled nanophotonic waveguides showed that nontrivial correlations emerge in the QW dynamics of two identical photons as a consequence of 
Hanbury-Brown-Twiss interference \cite{Peruzzo1500}. 
In Ref.\,\cite{PhysRevA.86.011603}, Lahini {\it et al.} precisely analyzed how such correlations are modified in the presence of interactions between the {\it walkers}.  Using the Bose-Hubbard (BH) model as a basis, they revealed that the dynamical evolution of two {\it walkers} changes greatly depending on both the interaction strengths and the initial state. 
This study sheds light on another important role of QWs as a fundamental building block of quantum simulators for many-body dynamics \cite{Bloch:2012aa,RevModPhys.86.153}. 

Quite recently, the continuous-time QWs of two interacting particles were demonstrated using bosonic ultracold atoms in a one-dimensional (1D) optical lattice \cite{Preiss13032015}. 
In this experiment, the high controllability of interatomic interactions is a great advantage when we investigate the dependence of particle correlations on the interaction strengths. 
Furthermore, the advanced technique provided by a quantum gas microscope \cite{Sherson:2010aa,Weitenberg:2011aa} allows us to access directly the dynamics of QWs by resolving each atom over lattice sites \cite{Preiss13032015,Fukuhara:2013ab}. 
The measured data quantitatively agree with theoretical calculations based on the BH model. 
These features convince us that ultracold atoms can offer a promising platform on which we develop quantum simulations via multiparticle QWs.   

We further expect that ultracold atoms will advance the study of QWs to the unexplored region where {\it walkers} contain internal degrees of freedom. 
The atom manipulation technique currently provides us with the multicomponent many-body system referred to as spinor Bose gases \cite{PhysRevLett.80.2027,PhysRevLett.81.742,PhysRevLett.87.010404,Kawaguchi2012253,RevModPhys.85.1191}. 
It is known that this system exhibits diverse and complex quantum phases caused by the interplay between interactions and spin degrees of freedom \cite{PhysRevLett.88.163001,PhysRevA.68.063602,PhysRevB.69.094410,PhysRevA.70.043628,PhysRevA.70.063610,PhysRevLett.94.110403,PhysRevLett.95.240404,PhysRevA.74.035601,PhysRevA.76.023606,Toga:2012aa,PhysRevB.88.104509}. 
In particular, the spin-1 bosonic atom system has been intensively studied as the simplest spinor Bose gas. The spin-dependent interaction of spin-1 atoms generates transitions among the 
spin states that preserve the $z$-component of the total spin \cite{PhysRevLett.81.5257}. 
This phenomenon is called spin-mixing dynamics and has been observed using a spin-1 Bose-Einstein condensate in a single optical trap  \cite{Stenger:1998aa,Chang:2005aa,PhysRevA.72.063619}  and also in an optical lattice  \cite{PhysRevLett.95.190405,1367-2630-8-8-152,PhysRevA.73.041602,PhysRevLett.114.225302}. 
Therefore, the QWs of spin-1 bosons present an intriguing problem, namely the clarification of the dynamical evolution of {\it walkers} that are interfering and interacting, and 
mixing spins under a condition where the total energy and total spins are both conserved. 

In this paper, we study a continuous-time QW including two spin-1 bosons trapped in a 1D optical lattice. 
We focus mainly on spin-mixing dynamics, which is one of the most intriguing features of spin-1 systems.
Furthermore, spin correlations as well as spatial correlations~\cite{PhysRevA.86.011603} can be studied with this model. Exploring the evolution of spin correlations helps towards an understanding of the dynamics involving spins in spin-1 lattice systems.


This paper is organized as follows. In Sec.~\ref{sec:model}, we introduce the spin-1 BH model and explain the spin-mixing dynamics in a single-site system. In Sec.~\ref{sec:exact}, we derive the exact solution of the two-particle dynamics governed by the spin-1 BH model. Using the results in Sec.~\ref{sec:exact}, we discuss the spin-mixing dynamics in quantum walks in Sec.~\ref{sec:mixing}. The dependence on the interaction strength is discussed in detail. In Sec.~\ref{sec:LRSC}, we explain how the spin-dependent interaction affects the evolution of spin correlations. Finally, we conclude this paper in Sec.~\ref{sec:conclusion}. In Appendix.~\ref{app:EH}, we derive the spin-mixing dynamics in an alternative way based on the effective Hamiltonian.

\section{\label{sec:model}Model}
We consider spin-1 bosons in a 1D optical lattice. These atoms are well described by the spin-1 BH Hamiltonian,
\begin{align}
\hat{H}=\hat{H}_J&+\hat{H}_{U_0}+\hat{H}_{U_2},\label{eq:hamiltonian}\\
\hat{H}_{J}&=-J\sum_{i,\alpha=0,\pm1}(\hat{b}_{i+1,\alpha}^{\dagger}\hat{b}_{i,\alpha}+\text{h.c.}),\label{eq:hopping}\\
\hat{H}_{U_0}&=\frac{U_0}{2}\sum_{i}\hat{n}_{i}(\hat{n}_i-1),\\
\hat{H}_{U_2}&=\frac{U_2}{2}\sum_{i}(\hat{\bf F}_i^2-2\hat{n}_i),
\end{align}
where $\hat{b}_{i,\alpha}^{\dagger} (\hat{b}_{i,\alpha})$ is the bosonic creation (annihilation) operator at the $i$-th site with the hyperfine spin state $\alpha~(=0,\pm1)$, $\hat{n}_i=\sum_{\alpha=0,\pm1}\hat{b}^{\dagger}_{i,\alpha}\hat{b}_{i,\alpha}$ is the corresponding local number operator. 
$\hat{\bf F}_i$ denotes the hyperfine spin operator at the $i$-th site defined 
in terms of $3 \times 3$ spin-1 matrices $F^{x, y, z}$ such as $\hat F_{i}^x=\sum_{\alpha, \beta} \hat b_{i,\alpha}^{\dag}(F^x)_{\alpha,\beta} \hat b_{i,\beta}$, etc.  $\hat{H}_J$, $\hat{H}_{U_0}$ and $\hat{H}_{U_2}$ represent the nearest neighbor hopping, spin-independent interactions and spin-dependent interactions, respectively. $\hat{H}_{U_2}$ induces a transition among states, which preserves the z-component of the total spin $\sum_{i}\hat{F}_i^z$.
The spin-dependence of the interactions arises from the difference between the scattering lengths for the total spins $F=0$ and $F=2$. To obtain an exact analysis, we restrict the discussion to 
a two-particle system. Furthermore, we set $\hbar$ and the lattice constant at unity throughout this paper. 

It is useful in relation to our later discussions on quantum walks that we briefly explain the dynamics of interacting two spin-1 bosons localized in a certain single site with only $\hat{H}_{U_0}$ and $\hat{H}_{U_2}$ (i.e., $J=0$). 
In the absence of an external magnetic field, we can discuss the intriguing spin-mixing dynamics without loss of generality in a limited case where the quantum state of spin-1 bosons 
is given by the superposition of two spin states with the same z-component of the total spin, i.e., $m_F=0$. Such states are  $|F=0,m_F=0\rangle=((\hat{b}^{\dagger}_0)^2-2\hat{b}^{\dagger}_{1}\hat{b}^{\dagger}_{-1})|0\rangle/\sqrt{6}$ and $|F=2,m_F=0\rangle=((\hat{b}^{\dagger}_0)^2+\hat{b}^{\dagger}_{1}\hat{b}^{\dagger}_{-1})|0\rangle/\sqrt{3}$, and the corresponding eigenenergies are 
$E_{F=0}=-2U_2$ and $E_{F=2}=U_2$, respectively \cite{PhysRevLett.84.1066,PhysRevLett.84.4031}. The time evolution of the quantum mechanical average of an operator $\hat{O}$ 
is evaluated via the state $|\psi(t)\rangle$ at time $t$: 
\begin{align}
\langle\hat{O}\rangle_t&=\langle\psi(t)|\hat{O}|\psi(t)\rangle, \nonumber\\
&=\sum_{F,F'}e^{i(E_F-E_{F'})t}\sum_{m_F,m_F'}[\langle F,m_F|\hat{O}|F',m_F'\rangle\nonumber\\
&\hspace{5em}\times\langle\psi(0)|F,m_F\rangle\langle F',m_F'|\psi(0)\rangle].\label{eq:obs}
\end{align}

\begin{figure}
\includegraphics[width=8cm]{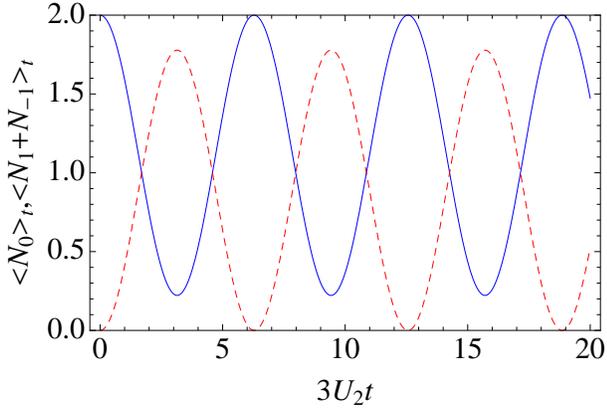}
\caption{
\label{fig:fig8}(Color online) Spin-mixing dynamics in the single site system ($J=0$). The solid and dashed lines represent the populations as a function of time in the spin state $\alpha=0$ and the sum of the populations in the spin states $\alpha=\pm1$, respectively.
}
\end{figure}

Let us consider a case where two atoms stay in the hyperfine spin state of $\alpha=0$ at the initial time $t=0$: $|\psi(0)\rangle = (\hat b_0^{\dag})^2/\sqrt{2}|0\rangle =\sqrt{1/3}|F=0,m_F=0\rangle+\sqrt{2/3}\,|F=2,m_F=0\rangle$. Using \Eqref{eq:obs}, we obtain the average number of atoms in the hyperfine spin state $\alpha=0,\pm1$ at time $t$:
\begin{align}
\langle\hat{N}_0\rangle_t&=\frac{10+8\cos(3U_2t)}{9}.\label{eq:SimpleSpinMixing0}\\
\langle\hat{N}_1\rangle_t&=\frac{4-4\cos(3U_2t)}{9}\nonumber\\
&=\langle\hat{N}_{-1}\rangle_t \label{eq:SimpleSpinMixing1}
\end{align}
Figure~\ref{fig:fig8} shows the time evolution of the spin-state populations calculated from Eqs.~\eqref{eq:SimpleSpinMixing0} and \eqref{eq:SimpleSpinMixing1}. The spin-mixing dynamics emerges owing 
to the $\hat{H}_{U_2}$ term in the Hamiltonian. The oscillation frequency $3U_2~(=E_{F=2}-E_{F=0})$ corresponds to the energy difference between the two states that we consider here. 

In the following sections, we show how such regular spin-mixing dynamics is modified by the inter-site hopping processes ($J\neq 0$).

\section{\label{sec:exact}Exact solution}
In this section, we analyze the quantum dynamics of two spin-1 bosons based on the Hamiltonian in \Eqref{eq:hamiltonian}. 
We derive the time-dependent wave function exactly by employing the method developed by A. Deuchert {\it et al.} in Ref.\,\cite{PhysRevA.86.013618}. 

\begin{figure}
\includegraphics[width=8cm]{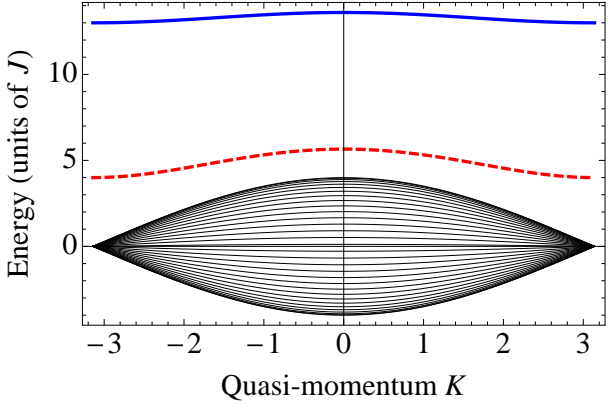}
\caption{
(Color online) Energy spectra for $U_2/U_0=0.3$ and $U_0/J=10$ as a function of center-of-mass quasi-momentum $K$. The thick solid and dashed lines correspond to the energy bands for bound states with $F=2$ and $F=0$, respectively. The bundle of thin solid lines represents the scattering continuum. The energy is defined in units of $J$.}
\label{fig:energy_band}
\end{figure}

The hopping of spin-1 atoms does not change their internal spin states, i.e., the two states $|F=0, m_F=0\rangle$ and $|F=2, m_F=0\rangle$ are not connected to each other via the hopping process. This allows us to straightforwardly generalize the bases $|F,m_F\rangle$ in the single-site system to the bases in the lattice system:
\begin{align}
|\psi_{-2U_2}\rangle_{i,j}&=\frac{1}{\sqrt{6}}\left[\hat{b}^{\dagger}_{i,0}\hat{b}^{\dagger}_{j,0}-\hat{b}^{\dagger}_{i,1}\hat{b}^{\dagger}_{j,-1}-\hat{b}^{\dagger}_{i,-1}\hat{b}^{\dagger}_{j,1}\right]|0\rangle,\label{eq:basis1}\\
|\psi_{U_2}\rangle_{i,j}&=\frac{1}{\sqrt{3}}\left[\hat{b}^{\dagger}_{i,0}\hat{b}^{\dagger}_{j,0}+\frac{\hat{b}^{\dagger}_{i,1}\hat{b}^{\dagger}_{j,-1}+\hat{b}^{\dagger}_{i,-1}\hat{b}^{\dagger}_{j,1}}{2}\right]|0\rangle. \label{eq:basis2}
\end{align}
Here the orthonormality is satisfied such that $_{i,j}\langle\psi_{\lambda}|\psi_{\lambda'}\rangle_{k,\ell}=\delta_{\lambda,\lambda'}(\delta_{i,k}\delta_{j,\ell}+\delta_{i,\ell}\delta_{j,k})/2$ where $\lambda$ and $\lambda'$ take $-2U_2$ or $U_2$. These bases also span all the eigenstates of the Hamiltonian in \Eqref{eq:hamiltonian}. Furthermore, in each spanned space represented by the quantum number $\lambda=-2U_2$ or $U_2$,  the Hamiltonian becomes equivalent to the spinless BH Hamiltonian  $\hat{H}_J+\hat{H}_{U_0}$ by replacing $U_0$ with $U_0-2U_2$ for \Eqref{eq:basis1} and with $U_0+U_2$ for \Eqref{eq:basis2}. This means that the dynamics of two interacting spin-1 bosons is essentially 
identical to that of spinless bosons, which greatly simplifies the theoretical treatment. 
Since the two-particle dynamics governed by the spinless BH model is exactly solvable by introducing center-of-mass coordinates $R=(i+j)/2$ and relative coordinates $r=i-j$ \cite{PhysRevA.86.013618}, we can calculate the exact dynamics for the spin-1 BH model. The eigenenergies and eigenstates in each space specified by $\lambda$ consist of scattering states and bound states. Hence the Schr\"{o}dinger equations are written as
\begin{align}
\hat{H}|\Psi_{\lambda,K}^B\rangle=E_{\lambda,K}^B|\Psi_{\lambda,K}^B\rangle,\nonumber\\
\hat{H}|\Psi_{\lambda,K,k}^S\rangle=E_{K,k}^S|\Psi_{\lambda,K,k}^S\rangle,
\end{align}
where $K$ and $k$ represent the center-of-mass and the relative quasi-momenta, respectively. We obtain the eigenenergies and eigenstates  
\begin{widetext}
\begin{align}
E^B_{\lambda,K}&=\sign(U_0+\lambda)\sqrt{(U_0+\lambda)^2+16J^2 [\cos(K/2)]^2},\\
|\psi^B_{\lambda,K}\rangle&=\sum_{R,r}\hspace{0em}'\frac{1}{\sqrt{2\pi}}e^{iKR}\frac{\sqrt{|{\cal U}_{\lambda,K}|}}{({\cal U}_{\lambda,K}^2+1)^{1/4}}
\left[{\cal U}_{\lambda,K}-{\rm sign}(U_0+\lambda)\sqrt{{\cal U}_{\lambda,K}^2+1}\right]^{|r|}
|\psi_{\lambda}\rangle_{R+r/2,R-r/2},
\end{align}
for bound states and
\begin{align}
E^S_{K,k}&=-4J\cos(K/2)\cos(k),\label{eq:Es}\\
|\psi^S_{\lambda,K,k}\rangle&=\sum_{R,r}\hspace{0em}'\frac{\frac{1}{\sqrt{2\pi}}e^{iKR}}{\sqrt{\pi\left(1+\frac{{\cal U}_{\lambda,K}^2}{\sin^2(k)}\right)}}\left[\cos(kr)+\frac{{\cal U}_{\lambda,K}}{\sin(k)}\sin(k|r|)\right]|\psi_{\lambda}\rangle_{R+r/2,R-r/2},\label{eq:psi_s}
\end{align}
\end{widetext}
for scattering states. Here we employ the abbreviations: $J_K=2J\cos(K/2)$, ${\cal U}_{\lambda,K}=(U_0+\lambda)/2J_K$ and $\sum_{R,r}\hspace{0em}^{'}=\sum_{R\in{\bf Z}}\sum_{r\in2{\bf Z}}+\sum_{R\in{\bf Z}+1/2}\sum_{r\in2{\bf Z}+1}$. Note that the energies of the scattering states are independent of interactions.
Figure~\ref{fig:energy_band} illustrates the energy spectra as a function of center-of-mass quasi-momentum $K$. The band of bound states is split into two 
depending on the spin-dependent interaction $U_2$ and located above the continuum of scattering states when $U_0/J>0$ and $U_0>2U_2$. If we take $U_2/U_0=1/2$ or $-1$, one of the bands disappears.

Eigenstates satisfy the following orthonormality relations:
\begin{align}
\langle\psi^B_{\lambda',K'}|\psi^B_{\lambda,K}\rangle&=\delta_{\lambda,\lambda'}\delta(K-K'),\label{eq:ortho1}\\
\langle\psi^S_{\lambda',K',k'}|\psi^S_{\lambda,K,k}\rangle&=\delta_{\lambda,\lambda'}\delta(K-K')\delta(k-k'),\label{eq:ortho2}\\
\langle\psi^S_{\lambda',K',k'}|\psi^B_{\lambda,K}\rangle&=0.\label{eq:ortho3}
\end{align}

\begin{figure}
\includegraphics[width=8cm]{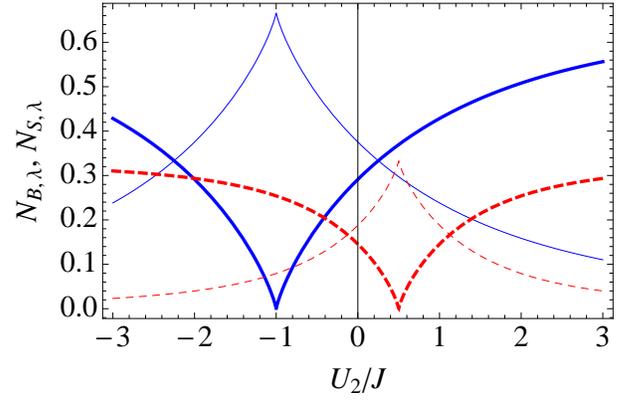}
\caption{
(Color online) The normalized number of bound states and scattering states in each space spanned by Eqs.~\eqref{eq:basis1} and \eqref{eq:basis2} as a function of the spin dependent interaction. The thick and thin lines represent the normalized number of atoms in the  bound states and scattering states, respectively. The solid and dashed lines correspond to the indices $\lambda=U_2$ and $\lambda=-2U_2$, respectively. We assume a condition where two $\alpha=0$ atoms occupy the same site and we choose $U_0/J=1$.}
\label{fig:bsr}
\end{figure}

Now the initial state at time $t = 0$ is generally written as a superposition of the eigenstates Eqs. \eqref{eq:basis1} and \eqref{eq:basis2},
\begin{align}
|\Psi(0)\rangle&=\sum_{\lambda}\int^{\pi}_{-\pi}dK\left[a_{\lambda,K}|\psi^B_{\lambda,K}\rangle+\int^{\pi}_0dk~b_{\lambda,K,k}|\psi^S_{\lambda,K,k}\rangle\right],
\end{align}
where the coefficients $a_{\lambda,K}$ and $b_{\lambda,K,k}$ satisfy the proper normalization condition
\begin{align}
\langle\Psi(0)|\Psi(0)\rangle&=\sum_{\lambda}\int^{\pi}_{-\pi}dK\left[|a_{\lambda,K}|^2+\int^{\pi}_0dk~|b_{\lambda,K,k}|^2\right]\nonumber\\
&=1.\label{eq:norm}
\end{align}
When two atoms are initially located at the same site, the normalized number of atoms in the bound states of the subspace $\lambda$(first term of \Eqref{eq:norm}) becomes
\begin{align}
N_{B,\lambda}&=\int^{\pi}_{-\pi}dK~|a_{\lambda,K}|^2\nonumber\\
&=c_{\lambda}\frac{2}{\pi}\frac{U_0+\lambda}{E^B_{\lambda,0}}\,G\left(\frac{16J^2}{E^B_{\lambda,0}\hspace{0em}^2}\right).\label{eq:nb}
\end{align}
Here,  $G(m)=\int_0^{\pi/2}\frac{1}{\sqrt{1-m\sin^2\theta}} d\theta$ represents the complete elliptic function of the first kind and $c_{\lambda}$ is the normalized number of atoms in the subspace $\lambda$,  which is determined by the choice of the initial state. If we start from two $\alpha=0$ atoms at the same site, $c_{U_2}=2/3$ and $c_{-2U_2}=1/3$. 
Figure~\ref{fig:bsr} shows the normalized number of bound states and scattering states in each space with respect to the spin-dependent interaction. Because of \Eqref{eq:norm}, the normalized number of atoms in the scattering states of the subspace $\lambda$ becomes $N_{S,\lambda}=c_{\lambda}-N_{B,\lambda}$. 
Clearly, the normalized number of atoms in the bound states, \Eqref{eq:nb}, increases with the absolute value of the interaction in each space. This is natural because the bound states are created by the interaction.

By definition, the state at time $t$ is straightforwardly given by
\begin{align}
|\Psi(t)\rangle&=\sum_{\lambda}\int^{\pi}_{-\pi}dK\Bigl[a_{\lambda,K}e^{iE^B_{\lambda,K}t}|\psi^B_{\lambda,K}\rangle\Bigr.\nonumber\\
&\hspace{7em}\Bigl.+\int^{\pi}_0dk~b_{\lambda,K,k}e^{iE^S_{K,k}t}|\psi^S_{\lambda,K,k}\rangle\Bigr]\nonumber\\
&\equiv \sum_{i,j}\sum_{\lambda}\Psi_{\lambda}(i,j,t)|\psi_{\lambda}\rangle_{i,j}.\label{eq:state}
\end{align}

This result enables us to calculate any physical quantities of the two-particle dynamics governed by the spin-1 BH model.

\section{\label{sec:mixing}spin-mixing dynamics in quantum walks}
In this section,  we discuss the spin-mixing dynamics in quantum walks. We show in detail that the inter-site hopping of atoms in the lattice system greatly changes the simple oscillation behavior of spin-mixing dynamics discussed in Sec.~\ref{sec:model}.

\subsection{Analytical results}

The spin-mixing dynamics is described by the total number of atoms in a hyperfine state $\alpha$. The quantum mechanical average at time $t$ ($\langle\bullet\rangle_t=\langle\Psi(t)|\bullet|\Psi(t)\rangle$) of the corresponding operator $\hat{N}_{\alpha}=\sum_i\hat{b}^{\dagger}_{i,\alpha}\hat{b}_{i,\alpha}$ is calculated to be 
\begin{align}
\langle\hat{N}_0\rangle_t
&=\frac{2}{3}\sum_{i,j}\Big(
|\Psi_{-2U_2}(i,j,t)|^2 +2|\Psi_{U_2}(i,j,t)|^2\nonumber\\
&~+2\sqrt{2}\re\left[\Psi^*_{-2U_2}(i,j,t)\Psi_{U_2}(i,j,t)\right]
\Big),\\
\langle\hat{N}_1\rangle_t
&=\frac{1}{3}\sum_{i,j}\Big(
2|\Psi_{-2U_2}(i,j,t)|^2 +|\Psi_{U_2}(i,j,t)|^2\nonumber\\
&~-2\sqrt{2}\re\left[\Psi^*_{-2U_2}(i,j,t)\Psi_{U_2}(i,j,t)\right]
\Big)\nonumber\\
&=\langle\hat{N}_{-1}\rangle_t.
\end{align}

The function defined in \Eqref{eq:state} can be expressed using the initial state $|\psi(0)\rangle$:
\begin{align}
\Psi_{\lambda}(i,j,t)&\equiv \sum_{i',j'}\hspace{0em}_{i',j'}\langle\psi_{\lambda}|\Psi(0)\rangle W^{\lambda}_{i,j;i',j'}(t),
\end{align}
where
\begin{align}
W^{\lambda}_{R,r;R',r'}(t)=\int^{\pi}_{-\pi}\frac{dK}{2\pi}e^{iK(R-R')}&\biglb[w^{B}_{\lambda,K}(r,r',t)\bigrb.\nonumber\\
&\biglb.+w^{S}_{\lambda,K}(r,r',t)\bigrb],\label{eq:melment}
\end{align}
is a matrix element of the time evolution operator in the space $\lambda$. Here, we introduce $R'=(i'+j')/2$ and $r'=i'-j'$. 
$w^{B}_{\lambda,K}(r,r',t)$ and $w^{S}_{\lambda,K}(r,r',t)$ correspond to the contributions of bound states and scattering states, respectively. The explicit formulae of these functions are
\begin{align}
w^{B}_{\lambda,K}(r,r',t)&=\frac{|{\cal U}_{\lambda,K}|}{\sqrt{1+{\cal U}_{\lambda,K}^2}}e^{-iE_{\lambda,K}^Bt}\nonumber\\
&\hspace{.5em}\times\left[{\cal U}_{\lambda,K}-\sign(U_0+\lambda)\sqrt{{\cal U}_{\lambda,K}^2+1}\right]^{|r|+|r'|},\label{eq:bcont}\\
w^{S}_{\lambda,K}(r,r',t)&=\int^{\pi}_{0}\frac{dk}{\pi}e^{-iE_{K,k}^St}\frac{f_{\lambda,K}(r)f_{\lambda,K}(r')}{1+\frac{{\cal U}_{\lambda,K}^2}{\sin^2(k)}},\label{eq:scont}
\end{align}
with
\begin{align}
f_{\lambda,K}(n)&=\cos(kn)+\frac{{\cal U}_{\lambda,K}}{\sin(k)}\sin(k|n|).
\end{align}

When two atoms in the hyperfine state $\alpha=0$ are initially located at the origin of a one-dimensional lattice, the projection of this initial state onto each space is given by
\begin{align}
\hspace{0em}_{i',j'}\langle\psi_{U_2}|\Psi(0)\rangle&=\sqrt{\frac{2}{3}}\delta_{i',0}\delta_{j',0},\\
\hspace{0em}_{i',j'}\langle\psi_{-2U_2}|\Psi(0)\rangle&=\frac{1}{\sqrt{3}}\delta_{i',0}\delta_{j',0}.
\end{align}
Hence the total number of $\alpha=0$ atoms at time $t$ becomes
\begin{align}
\langle\hat{N}_0\rangle_t
&=\frac{1}{9}\left(
10+8\sum_{R,r}\hspace{0em}^{'}\re\left[W^{-2U_2}_{R,r;0,0}\hspace{0em}^*(t)W^{U_2}_{R,r;0,0}(t)\right]
\right)\nonumber\\
&\equiv\frac{1}{9}\left[10+8(X_B(t)+X_S+X_{BS}(t))\right].\label{eq:n0}
\end{align}
From \Eqref{eq:melment}, the matrix element $W$ is the sum of the contributions of the bound states (\Eqref{eq:bcont}) and scattering states (\Eqref{eq:scont}). 
We can thus separate the time dependent term of $\langle\hat{N}_0\rangle_t$ into three parts: the product of the contribution of the bound states, $X_B$, the product of the contribution of the scattering states, $X_S$, and  the interference between the contributions of the bound and scattering states, $X_{BS}$.
Specifically,
\begin{widetext}
\begin{align}
X_B(t)&=\frac{(U_0-2U_2)(U_0+U_2)}{2U_0-U_2}\int^{\pi}_{-\pi}\frac{dK}{2\pi}\frac{E_{-2U_2,K}^B+E_{U_2,K}^B}{E_{-2U_2,K}^BE_{U_2,K}^B}\nonumber\\
&\hspace{8em}\times\cos\left[(E_{-2U_2,K}^B-E_{U_2,K}^B)t\right],\label{eq:X_B}\\
X_S&=1-\frac{2}{\pi}\frac{1}{2U_0-U_2}\left[\frac{(U_0+U_2)^2}{E_{U_2,0}^B}\,G\left(\frac{16J^2}{E_{U_2,0}^B\hspace{0em}^2}\right)+\frac{(U_0-2U_2)^2}{E_{-2U_2,0}^B}\,G\left(\frac{16J^2}{E_{-2U_2,0}^B\hspace{0em}^2}\right)\right],\\
X_{BS}(t)&=3U_2\int^{\pi}_{-\pi}\frac{dK}{2\pi}\int^{\pi}_{0}\frac{dk}{\pi}~\left[\frac{A_{K,k}(U_0-2U_2)}{1+\frac{{\cal U}_{U_2,K}^2}{\sin^2(k)}}-\frac{A_{K,k}(U_0+U_2)}{1+\frac{{\cal U}_{-2U_2,K}^2}{\sin^2(k)}}\right],\label{eq:X_BS}
\end{align}
with
\begin{align}
A_{K,k}(\lambda)&=\frac{U_0+\lambda}{E_{\lambda,K}^B}\frac{\left(U_0+\lambda-E_{\lambda,K}^B\right)\cos\left[\left(E_{\lambda,K}^B-E^S_{K,k}\right)t\right]}{16J^2[\cos(K/2)]^2+(U_0+\lambda+E^S_{K,k})\left(U_0+\lambda-E_{\lambda,K}^B\right)}. 
\end{align}
\end{widetext}
Here the function $G(m)$ represents the complete elliptic integral of the first kind. 
The time independent nature of $X_S$ comes from the interaction  independence of the energy of the scattering states.
$X_S$ is calculated via the product of $w_{-2U_2,K}^*$ and $w_{U_2,K}$, and its time dependence is determined by the difference between the energies in the exponential part included in the 
$w_{\lambda,K}$ function. 
However, $E^S_{K,k}$ in \Eqref{eq:Es} clearly shows that this energy difference vanishes and therefore $X_S$  becomes independent of time. 
Regarding $X_B$ and $X_{BS}$, it is difficult to derive their expressions as a function of time $t$ by analytically dealing with the integrals with respect to quasi-momenta in Eqs.~\eqref{eq:X_B} and \eqref{eq:X_BS}. 
Instead, at $t=0$, we can perform the integrals and obtain the following useful expressions: 
\begin{widetext}
\begin{align}
X_B(0)&=\frac{2}{\pi}\frac{(U_0+U_2)(U_0-2U_2)}{2U_0-U_2}\left[\frac{1}{E^B_{U_2,0}}\,G\left(\frac{16J^2}{E^B_{U_2,0}\hspace{0em}^2}\right)+\frac{1}{E^B_{-2U_2,0}}\,G\left(\frac{16J^2}{E^B_{-2U_2,0}\hspace{0em}^2}\right)\right],\\
X_{BS}(0)&=-\frac{2}{\pi}\frac{3U_2}{2U_0-U_2}\left[\frac{U_0-2U_2}{E^B_{-2U_2,0}}\,G\left(\frac{16J^2}{E^B_{-2U_2,0}\hspace{0em}^2}\right)-\frac{U_0+U_2}{E^B_{U_2,0}}\,G\left(\frac{16J^2}{E^B_{U_2,0}\hspace{0em}^2}\right)\right].
\end{align}
\end{widetext}
Then we can immediately find $X_B(0)+X_S+X_{BS}(0)=1$, which is consistent with the choice of the initial condition.
A similar calculation shows
\begin{align}
\langle\hat{N}_1\rangle_t
&=\frac{1}{9}\left[4-4(X_B(t)+X_S+X_{BS}(t))\right]\nonumber\\
&=\langle\hat{N}_{-1}\rangle_t.\label{eq:n1}
\end{align}
We note that \Eqref{eq:n0} (\Eqref{eq:n1}) has the same form as \Eqref{eq:SimpleSpinMixing0} (\Eqref{eq:SimpleSpinMixing1}) because the constant term comes from the norm of 
the wave functions in each space, which does not change with time.

\subsection{Numerical results}
\begin{figure*}
\subfloat[$X_{B},X_S$]{
\includegraphics[width=8cm]{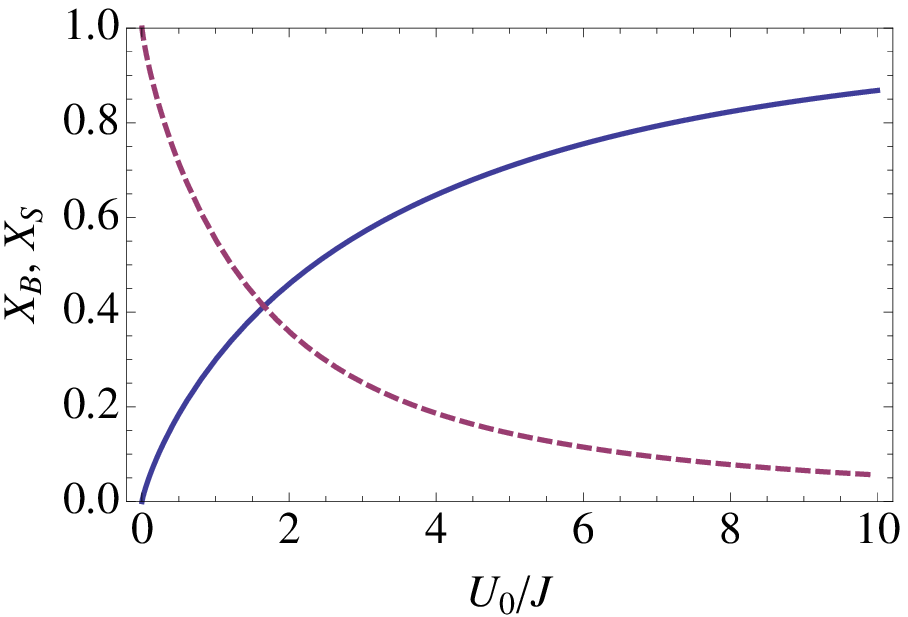}
\label{fig:fig9}}~
\subfloat[$X_{BS}$]{
\includegraphics[width=8cm]{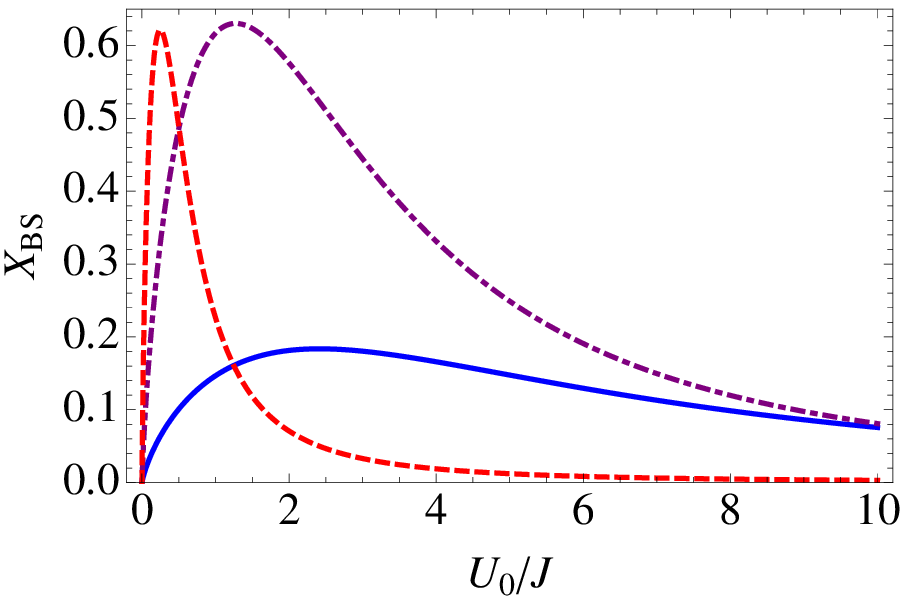}
\label{fig:fig10}
}
\caption{
(Color online) Interaction dependence of $X_B,~X_S$ and $X_{BS}$ at $t=0$. (a) $U_2/U_0=0.3$. The solid and dashed lines represent $X_B$ and $X_S$, respectively. (b) Interference term $X_{BS}(0)$ for three kinds of $U_2/U_0$ values. The solid, dot-dashed and dashed lines correspond to $U_2/U_0=0.3,~1$, and 5, respectively.}
\label{fig:int}
\end{figure*}

\begin{figure}
\includegraphics[width=8cm]{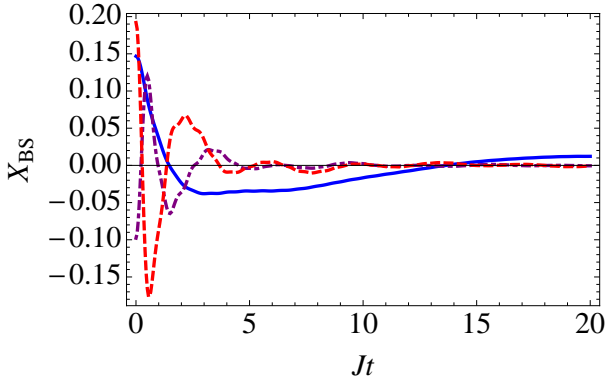}
\caption{
(Color online) Time dependence of the interference term $X_{BS}$. We choose three sets of interactions $U_0$ and $U_2$ in the vicinity of the maximum point of $X_{BS}(0)$ 
in Fig.~\ref{fig:fig10}. The solid, dot-dashed, and dashed lines correspond to $(U_0/J,~U_2/U_0)=(1,~0.3),~(2,~1)$, and $(0.5,~5)$, respectively.}
\label{fig:fig11}
\end{figure}

We carried out numerical calculations to reveal the properties of the spin-mixing dynamics in a lattice system. Although the ratio $U_2/U_0$ is rather small in experiments such as $^{23}$Na (positive) and $^{87}$Rb (negative), i.e., less than a few percent, here we choose $U_2/U_0=0.3$ to demonstrate the effect of spin-dependent interactions more clearly. 
Figure \ref{fig:int} shows $X_{B}$, $X_{S}$, and $X_{BS}$ at the initial time $t=0$ as a function of the normalized interaction strength $U_0/J$. 
In Fig. \ref{fig:fig9}, the contribution of the bound states $X_{B}(0)$ gradually increases with the interaction, while the contribution of the scattering states $X_{S}$ decreases with the interaction. This reflects the fact that the interaction reduces the number of particles in the scattering states (see \Eqref{eq:psi_s}). Since the energy of the initially localized state must be conserved, the interaction suppresses the dissociation of the pair \cite{Winkler2006RB}. On the other hand, the interference term $X_{BS}(0)$ exhibits a non-monotonic dependence on the interaction, reaching maximum at around $U_0/J\sim1$ (see Fig. \ref{fig:fig10}). However, $X_{BS}(t)$ rapidly decreases with time as shown in Fig. \ref{fig:fig11}. This characteristic time dependence can be interpreted by considering the evolution of the bound and scattering states. The wave functions of the scattering states spread over the lattice with time, while the wave functions of the bound states remain localized. The overlaps between these two kinds of states decrease with time. Hence we neglect $X_{BS}$ in the following discussion. 

\begin{figure*}
\subfloat[$U_0/J=20,~U_2/J=6$]{
\includegraphics[width=8cm]{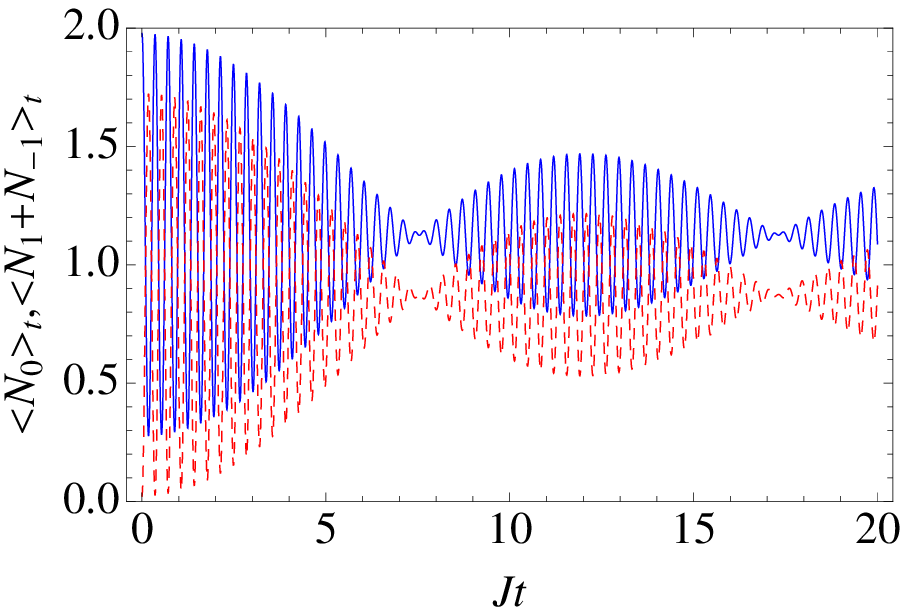}
}~
\subfloat[$U_0/J=10,~U_2/J=3$]{
\includegraphics[width=8cm]{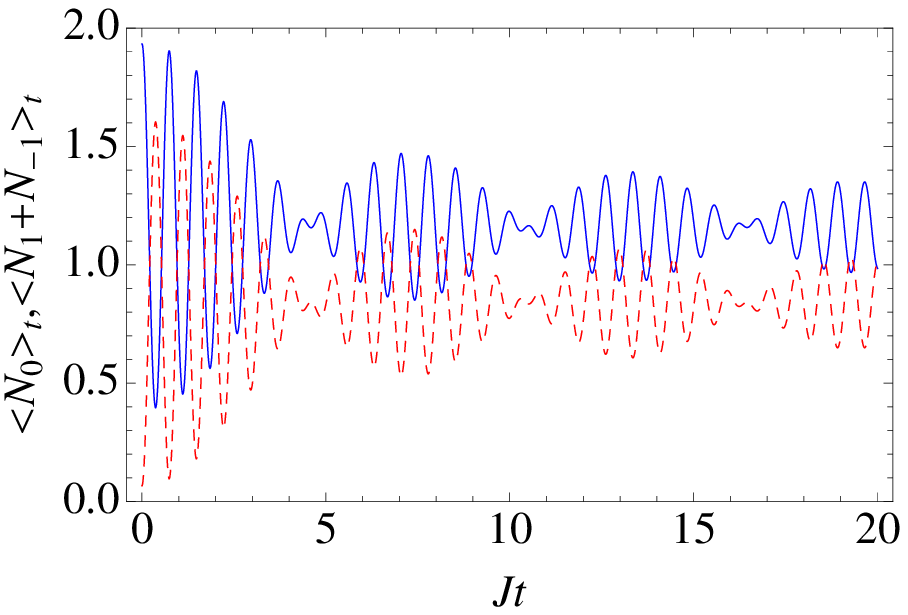}
}
\newline
\subfloat[$U_0/J=5,~U_2/J=1.5$]{
\includegraphics[width=8cm]{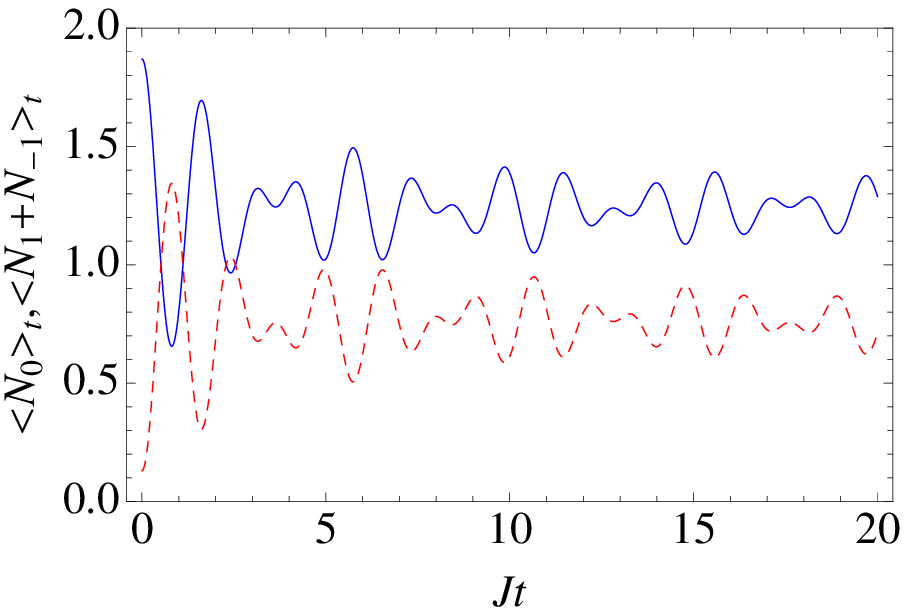}
}~
\subfloat[$U_0/J=1,~U_2/J=0.3$]{
\includegraphics[width=8cm]{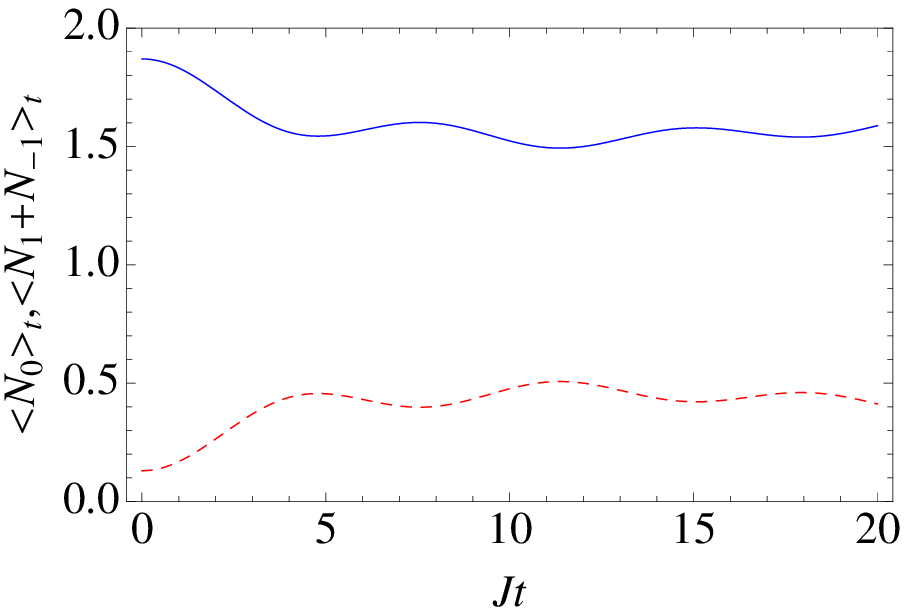}
}
\caption{
(Color online) Spin-mixing dynamics in a 1D optical lattice with $U _2/U_0 = 0.3$. The solid and the dashed lines represent the populations in the spin state $\alpha= 0$ and the sum of the populations in the spin states $\alpha=\pm1$, respectively.}
\label{fig:fig13}
\end{figure*}

Next we analyze the spin-mixing dynamics based on Eqs.~\eqref{eq:n0} and \eqref{eq:n1}. Figure \ref{fig:fig13} shows the time-evolution of the total number of atoms in the hyperfine state 
$\alpha$ corresponding to the four different $U_2/J$ values. We see that the spin-mixing dynamics is highly sensitive to the interactions. For a large $U_2/J$, there are two distinct frequencies and the amplitude of the slower oscillation gradually decreases with time. We elucidate the dependence of the two frequencies on the interactions from the results of a spectral analysis:  the higher frequency 
$\omega_{\rm high}$ coincides with the characteristic frequency of spin-mixing in the single site system $3U_2$ and the lower frequency $\omega_{\rm low}$ is reduced 
as the interaction decreases. For a small $U_2/J$ with the fixed ratio $U_2/U_0=0.3$, the spin-mixing dynamics is highly suppressed. This behavior comes from the fact that the coefficient of $X_B(t)$ becomes small in the vicinity of $U_0-2U_2=0$ or $U_0+U_2=0$. In these situations, the number of atoms in the bound states decreases. Moreover, the reduction of the frequencies $\omega_{\rm high}$ and $\omega_{\rm low}$ (see Fig.~\ref{fig:frequency}) around $U_2=0$ makes it difficult to observe the spin-mixing. Finally, all results discussed in this section are completely applicable 
when the $U_0$ sign changes while maintaining the ratio $U_2/U_0$, because of the symmetry of the dynamics governed by the 1D spinless BH model~\cite{PhysRevA.86.013618}.

\subsection{Discussions}\label{sec:smdisc}
Here, we reveal why two frequencies appear in the spin-mixing dynamics by taking the limits of both $U_0+U_2\gg4J$ and $U_0-2U_2\gg4J$. In these limits, $X_B(t)$ becomes
\begin{align}
X_{B}(t)&\simeq{\cal J}_0(2\epsilon t)\cos[(3U_2-2\epsilon)t]\label{eq:FB},
\end{align}
where ${\cal J}_n(x)$ denotes the Bessel function of the first kind. $\epsilon=J_{U_0-2U_2}-J_{U_0+U_2}$  is related to the cotunneling process, namely the simultaneous hopping of two particles at the same site to an adjacent site. $J_U\equiv2J^2/U$ is the effective hopping of the cotunneling process in the large interaction limit, $U/J\gg1$ \cite{Folling:2007aa}. Using the addition theorem of the Bessel function: ${\cal J}_{m}(x-y)=\sum_{n=-\infty}^{\infty}{\cal J}_n(x){\cal J}_{n-m}(y)$ with $|x|>|y|$, the factor ${\cal J}_0(2\epsilon t)$ can be rewritten as
\begin{align}
\nonumber {\cal J}_0(2\epsilon t)& =\sum_{n}{\cal J}_n(2J_{U_0-2U_2}t){\cal J}_{n}(2J_{U_0+U_2}t),\\ 
&~=\sum_n\psi_{J_{U_0-2U_2}}^*(n,t)\psi_{J_{U_0+U_2}}(n,t).
\end{align}
Here, $\psi_J(n,t)=i^{|n|}{\cal J}_{|n|}(2Jt)$ is the wave function of the continuous-time QW (dynamics of single particle initially located at the origin, governed by $H_J$) at the $n$-th site at time $t$ \cite{PhysRevE.72.026113}. Hence, one can say that the Bessel function in \Eqref{eq:FB} represents the overlap of the bound-state wave functions in different bands. 
In the limit of $U_0/J\to\infty$, $\epsilon$ becomes $0$ and thus the Bessel function becomes 1. Since $X_S$ and $X_{BS}$ disappear in this limit, \Eqref{eq:n0} (\Eqref{eq:n1}) coincides with \Eqref{eq:SimpleSpinMixing0} (\Eqref{eq:SimpleSpinMixing1}).
Note that \Eqref{eq:FB} can also be derived by using  the effective Hamiltonian for bound states (see appendix \ref{app:EH}). 

\begin{figure*}
\subfloat[]{
\includegraphics[width=8cm]{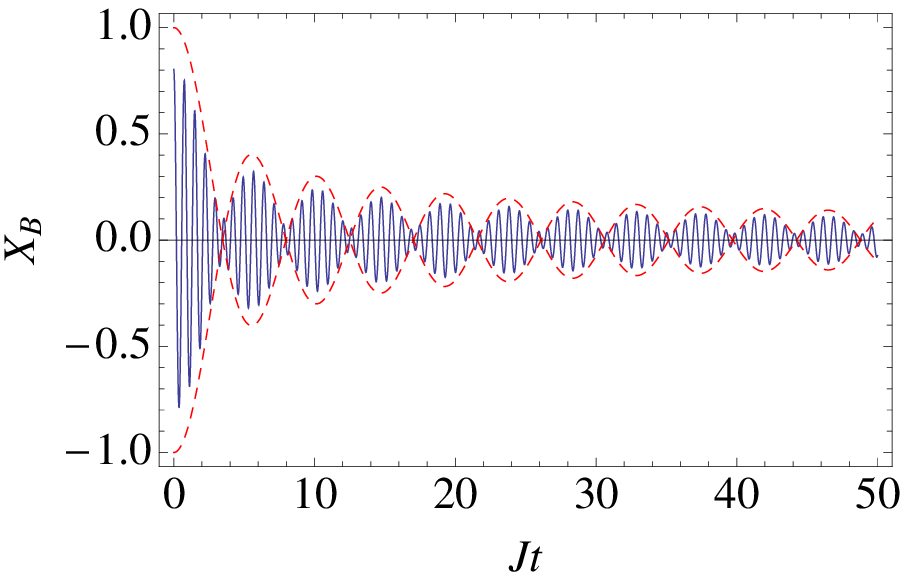}
\label{fig:envelop}
}~
\subfloat[]{
\includegraphics[width=8cm]{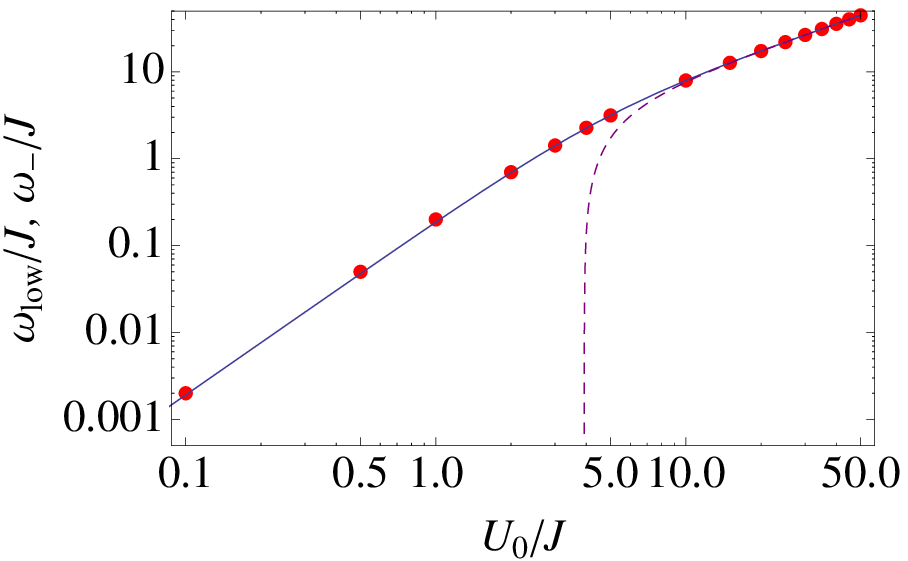}
\label{fig:frequency}
}
\caption{
(Color online) (a)  Time dependence of bound state term $X_B(t)$. Interactions are $U_0/J=10$, and $~U_2/U_0=0.3$. The envelope functions plotted by the dashed lines 
correspond to $\pm{\cal J}_0(\epsilon't)$. (b) Three characteristic frequencies of spin-mixing dynamics 
as a function of $U_0$: $\omega_{\rm low}$ (circle), $\omega_-(\epsilon)$ (dashed line) and $\omega_-(\epsilon')$ (solid line). }
\label{fig:fig12}
\end{figure*}

Since \Eqref{eq:FB} is the product of periodic and quasi-periodic functions, the frequencies of the spin-mixing dynamics are determined by the  sum and the difference between the frequencies of each function.  The sum $\omega_+=[(3U_2-2\epsilon)+2\epsilon]=3U_2$ is identical to the frequency in a single site system (see \Eqref{eq:SimpleSpinMixing0}, \eqref{eq:SimpleSpinMixing1}). Because the approximation in \Eqref{eq:FB} cannot be established for $U/J\sim1$, the difference $\omega_-(\epsilon)=[(3U_2-2\epsilon)-2\epsilon]=3U_2-4\epsilon$ does not coincide with $\omega_{\rm low}$, 
which is the smaller frequency calculated from the spin-mixing dynamics (see the dashed line and circles in Fig. \ref{fig:frequency}). Since $4J_U$ coincides with the bandwidth of the bound states in the large interaction limit, we consider the exact bandwidth of the bound states $4J_U'=U-\sign(U)\sqrt{U^2+16J^2}$, instead of $4J_U$. Then $\epsilon$ becomes 
\begin{align}
\epsilon'=|J_{U_0-2U_2}'-J_{U_0+U_2}'|.
\end{align}
Substituting $\epsilon'$ for $\omega_-$, $\omega_-(\epsilon')$ coincides with $\omega_{\rm low}$ (see the solid line and circles in Fig. \ref{fig:frequency}). As shown in Fig. \ref{fig:envelop}, $\pm{\cal J}_0(2\epsilon't)$ well describes the envelope function of $X_B(t)$.  Surprisingly, $\omega_+$ is always correct even for small interactions (compared with $\omega_{\rm high}$).

\section{\label{sec:LRSC}Emergence of Long-range Spin Correlations}
Quantum correlations in the 1D bosonic two-particle QW are discussed in Refs.~\cite{PhysRevA.86.011603,Preiss13032015} on the basis of the spinless BH model. 
It has been clarified that the time evolution of two-particle correlation depends strongly on both the interaction strength and the initial condition of the QWs. 
When two bosons are initially localized at the same site, the relative motion is suppressed with increases in interactions. In contrast, when two bosons are initially located at adjacent sites, the relative motion is enhanced as the interaction increases. These dynamical properties can be understood by noting the energy conservation of the system. 
Let us consider a case where two bosons are localized at the same site. 
The repulsive (attractive) interaction makes the energy of this boson pair higher (lower) than the energy of the other states. 
The two bosons therefore tend to maintain their localized states to conserve energy \cite{Winkler2006RB}. For the same reason, the spatially separated bosons rarely occupy the same site, 
leading to the enhancement of relative motion. 

In the spin-1 boson system, we observed the similar two-particle correlations mentioned above. Thus we focus on the evolution of 
two-particle spin correlations given by $\langle \hat{F}_{z,i}\hat{F}_{z,j}\rangle_t$. To elucidate the role of spin-dependent interaction, we choose the initial state, which does not have 
non-local spin correlations ($i\neq j$) for $U_2=0$. This state corresponds to the superposition of a parallel spin state and an anti-parallel spin state with an equal ratio: 
\begin{align}
|\Psi_{i,j}^{\theta}(0)\rangle=A_{i,j}\hat{B}^{\dagger}_i(\theta)\hat{B}^{\dagger}_j(\theta)|0\rangle, 
\end{align}
where $A_{i,j}\equiv1/(\sqrt{4+2\delta_{i,j}})$ is a normalization factor and $\hat{B}^{\dagger}_j(\theta)\equiv\hat{b}^{\dagger}_{j,1}+e^{i\theta}\hat{b}^{\dagger}_{j,-1}$ is a corresponding creation operator with 
an arbitrary angle $\theta$. 
From Eqs.~\eqref{eq:basis1} and \eqref{eq:basis2}, and also by introducing the spin states ~$|\hspace{-.25em}\uparrow\uparrow\rangle_{i,j}\equiv\hat{b}^{\dagger}_{i,1}\hat{b}^{\dagger}_{j,1}|0\rangle/\sqrt{2}$ and $|\hspace{-.25em}\downarrow\downarrow\rangle_{i,j}\equiv \hat{b}^{\dagger}_{i,-1}\hat{b}^{\dagger}_{j,-1}|0\rangle/\sqrt{2}$, we can rewrite the initial state in a more informative way:  
\begin{widetext}
\begin{align}
|\Psi_{i,j}^{\theta}(0)\rangle=\sqrt{2}A_{i,j}\left[|\hspace{-.25em}\uparrow\uparrow\rangle_{i,j}+e^{2i\theta}|\hspace{-.25em}\downarrow\downarrow\rangle_{i,j}
+\sqrt{\frac{2}{3}}e^{i\theta}|\psi_{U_2}\rangle_{i,j}-\frac{2}{\sqrt{3}}e^{i\theta}|\psi_{-2U_2}\rangle_{i,j}\right].
\end{align}
\end{widetext}
Here $|\hspace{-.25em}\uparrow\uparrow\rangle_{i,j},~|\hspace{-.25em}\downarrow\downarrow\rangle_{i,j}$ and $|\psi_{U_2}\rangle_{i,j}$ correspond to the three basis states 
of the total spin $F=2$ states with an interaction energy of $U_0+U_2$, and they give the positive spin correlations. On the other hand, $|\psi_{-2U_2}\rangle_{i,j}$ 
is the basis of the $F=0$ state with an interaction energy of $U_0-2U_2$, and it gives the negative spin correlations. 
All these four states evolve separately over time while keeping their spin states as discussed in Sec.\,\ref{sec:exact}. 
Therefore, the time dependence of two-particle spin correlations is determined by the quantum-mechanical superposition of spins during the dynamical evolution in QWs. 
Under the condition of a finite $U_2$, we can expect the emergence of non-local spin correlations owing to the difference in interaction energy mentioned above.  

\begin{figure*}
\subfloat[$U_2/U_0=0.3$]{
\includegraphics[width=8cm]{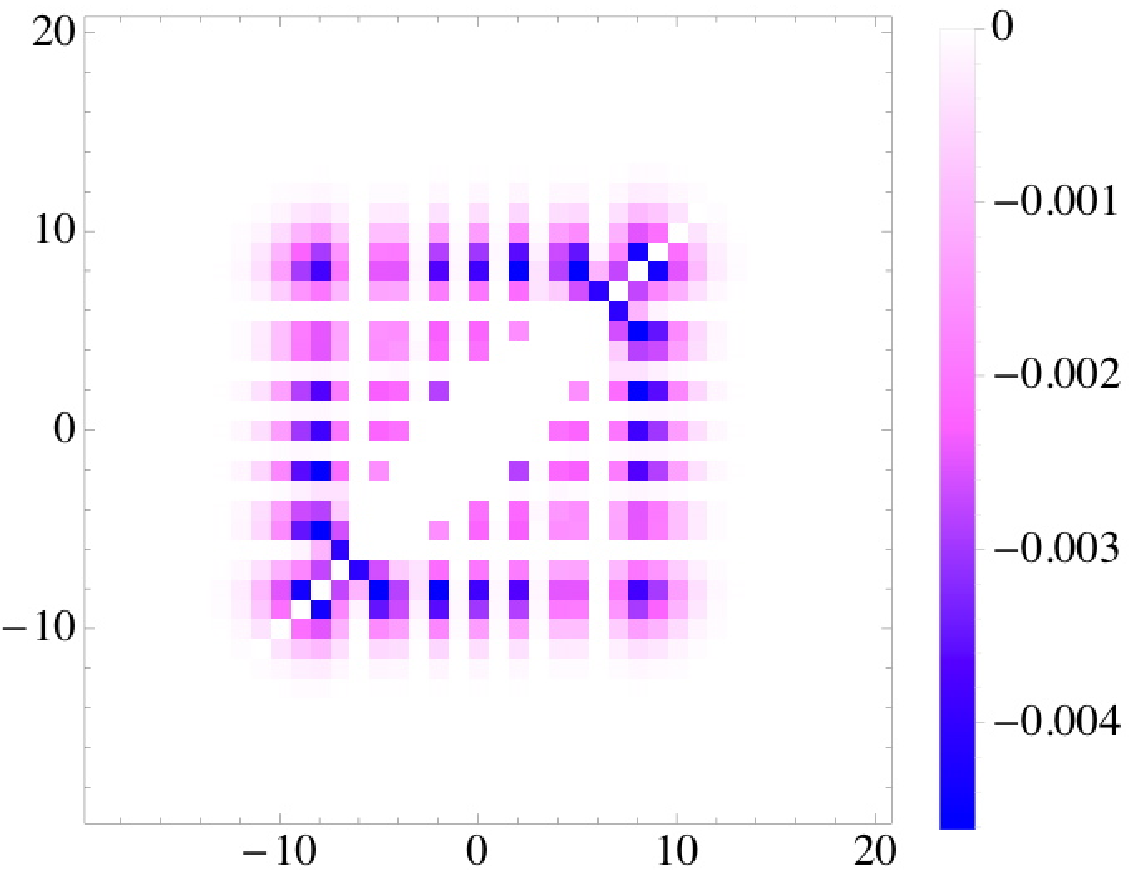}\label{fig:same+}
}~
\subfloat[$U_2/U_0=-0.3$]{
\includegraphics[width=8cm]{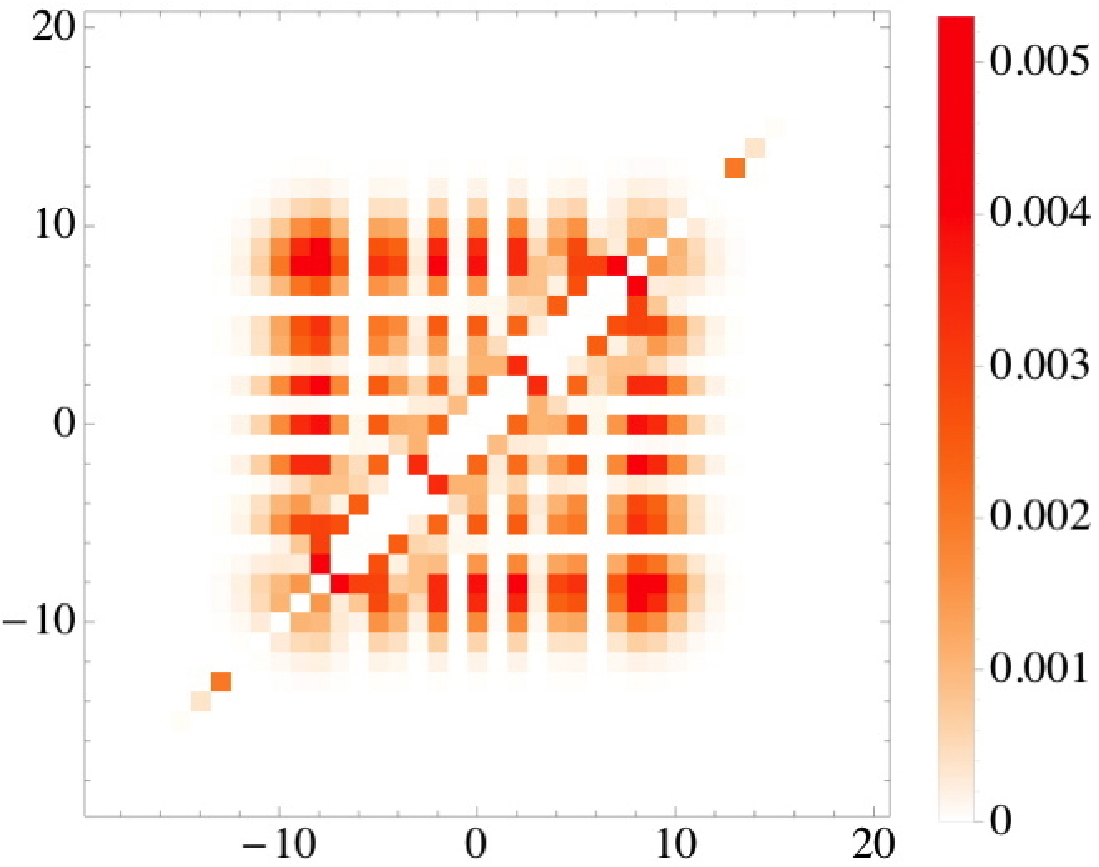}\label{fig:same-}
}\\
\subfloat[$U_2/U_0=0.3$]{
\includegraphics[width=8cm]{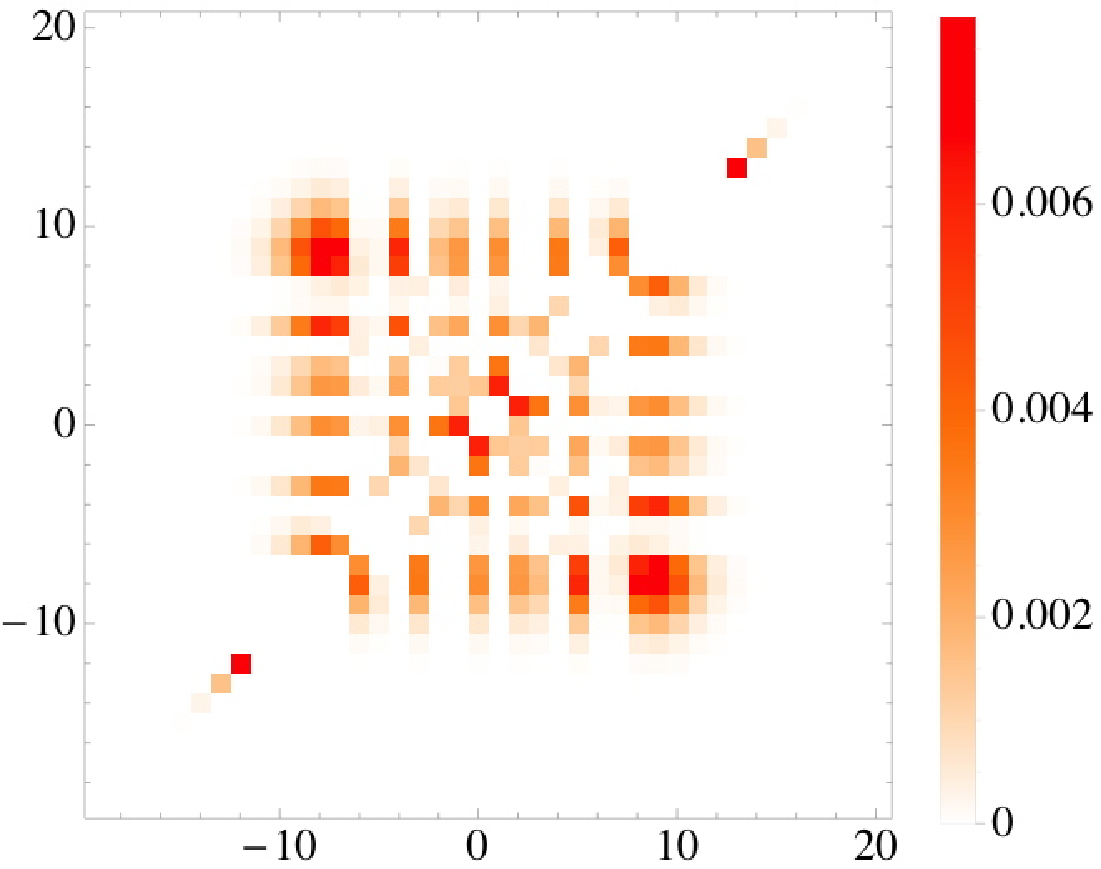}\label{fig:ad+}
}~
\subfloat[$U_2/U_0=-0.3$]{
\includegraphics[width=8cm]{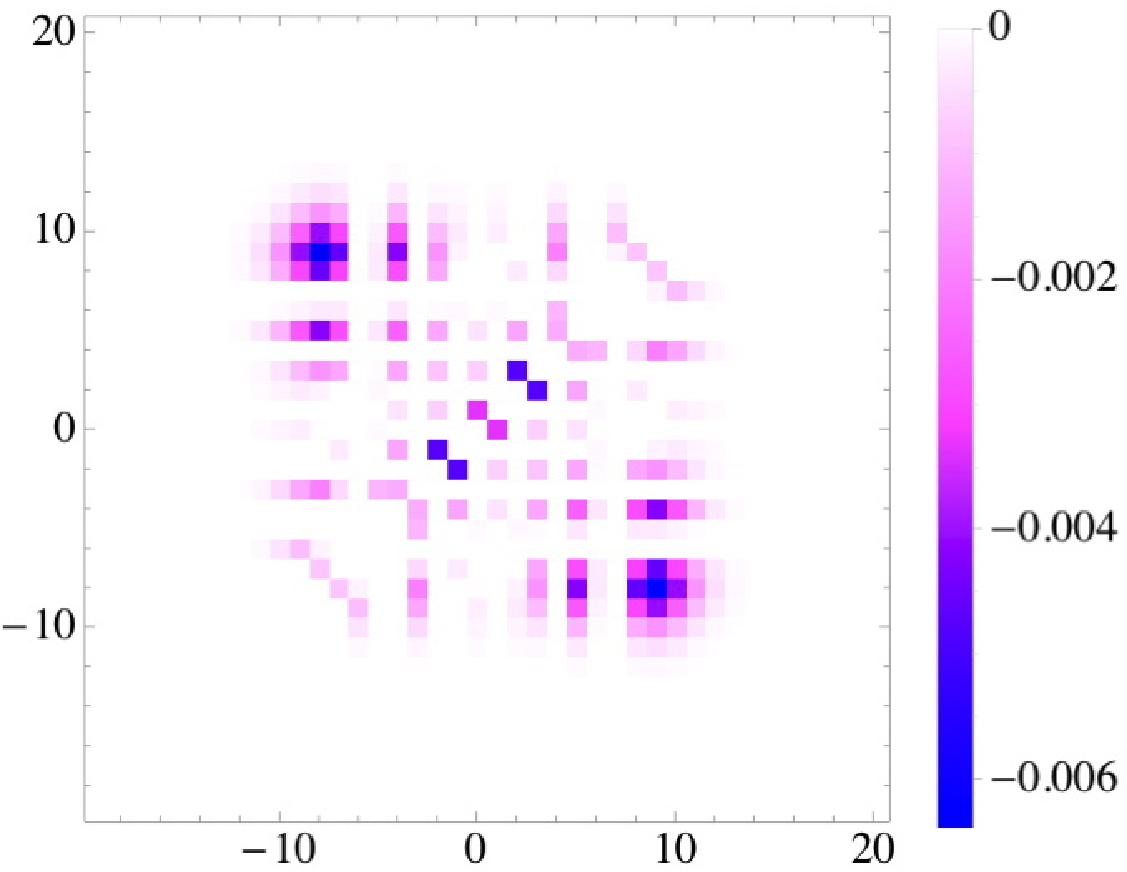}\label{fig:ad-}
}
\caption{
(Color online) Two-particle spin correlations at $t=5/J$ calculated with $U_0/J=2$ and $\theta=0$. The vertical and horizontal axes represent the lattice site indices. (a) and (b) Spin correlations starting from the initial state $|\Psi_{0,0}^{0}(0)\rangle$ where two spin-1 bosons are placed at the origin. (c) and (d) Spin correlations for the initial state $|\Psi_{0,1}^{0}(0)\rangle$ where two bosons are placed 
at the origin and the 1st site. We assume a positive $U_2$ in (a) and (c), and a negative $U_2$ in (b) and (d).}
\label{fig:fig02}
\end{figure*}

Figure \ref{fig:fig02} shows simulation results for spin correlations at $t=5/J$ calculated with $U_0/J=2$ and $\theta=0$. 
We further assume the spin-dependent interaction: $U_2=0.3U_0$ in Fig.\,\ref{fig:same+} and Fig.\,\ref{fig:ad+}; $U_2=-0.3U_0$ in Fig.\,\ref{fig:same-} and Fig.\,\ref{fig:ad-}. 
We find that the long-range spin correlations depend strongly on the sign of  $U_2$ and the initial states, 
which is a characteristic of two-particle QWs of interacting spin-1 bosons. 

First we consider a case where two spin-1 bosons are initially localized at the origin $|\Psi_{0,0}^{0}(0)\rangle$. 
Long-range spin correlations ($|i-j|\gg1$) are negative for $U_2/U_0>0$ (see Fig.\,\ref{fig:same+}) and positive for $U_2/U_0>0$ (see Fig.\,\ref{fig:same-}). 
We can understand this property as follows. 
The interaction greatly suppresses the relative motion of two spin-1 bosons for this initial state, which is similar to the spinless case. 
On the other hand, a spin-dependent interaction reduces the whole interaction energy and then enhances the relative motions for the $F=0$ ($F=2$) 
states when $U_2/U_0>0$ ($U_2/U_0<0$).  Correspondingly, long-range spin correlations become negative (positive). 

Next we start from the spatially separated initial state $|\Psi_{0,1}^{0}(0)\rangle$, i.e., each spin-1 boson is initially located at the origin and the 1st site. 
Long-range spin correlations become positive for $U_2/U_0>0$ (see Fig.\,\ref{fig:ad+}) and negative for $U_2/U_0<0$ (see Fig.\,\ref{fig:ad-}). For this type of initial state, the interaction enhances relative motions.  By noting the spin-dependent interaction, the relative motions of $F=2$ ($F=0$) states are relatively enhanced for $U_2/U_0>0$ ($U_2/U_0<0$), leading 
to positive (negative) long-range spin correlations.

Note that two-particle spin correlations do not show any dependence on the angle $\theta$. Therefore, we show the results for $\theta=0$ only. 

\section{\label{sec:conclusion}Conclusion}
In this work, we studied the QWs of interacting spin-1 bosons on the basis of the 1D spin-1 BH model. We derive an exact expression for the time-dependent wave function by extending the method developed in Ref.\,\cite{PhysRevA.86.013618} to a case including spin degrees of freedom. Using this expression, the spin-mixing dynamics in QWs is discussed in detail both analytically and numerically. 
We show that the spin-mixing dynamics is characterized by two frequencies in the limit of large spin-dependent interaction. 
One of the two frequencies is determined by the energy difference between two bound states and coincides with the characteristic frequency of the spin-mixing dynamics in a single site system. The other frequency is related to the cotunneling process that is the simultaneous hopping of a pair of atoms between lattice sites. These properties indicate that the dynamics in the spin space is strongly correlated 
to the dynamical evolution in real space via spin-dependent interactions. We find that the spin-mixing amplitude is suppressed in the vicinity of interactions satisfying $U_0-2U_2=0$ or $U_0+U_2=0$ because the number of spin-1 bosons in the bound states is greatly reduced there. 

We also numerically investigate two-particle spin correlations in the present system. Long-range spin-correlations emerge and the signs of the spin correlations can be controlled by changing the sign of spin-dependent interaction and/or the initial condition. This comes from the fact that the spin-dependent interaction effectively shifts the spin-independent interaction in accordance with the subspace specified by the total spin $F=0$ or $F=2$.


Experiments with ultracold atoms have been making rapid progress in recent years. The results presented here will be demonstrated experimentally in the near future. 
An interesting idea for future work is to extend the present study to a case including quadratic Zeeman effects, which are induced by magnetic fields \cite{PhysRev.38.2082.2} or microwaves \cite{PhysRevA.73.041602}, and examine how the spin-mixing dynamics is modified.
Although we focus on a system of interacting spin-1 bosons in this paper, generalization to other spinfull boson systems can be realized by performing similar calculations. On the other hand, from the viewpoint of two-particle dynamics, spin-mixing is a universal phenomenon in various spinfull systems except for spin-1/2. It might be another intriguing problem to study spin-mixing dynamics in fermionic systems  and 
reveal the qualitative difference between bosons and fermions. 

Our study paves the way for exploring continuous-time QWs including internal degrees of freedom. 
This opens up the possibility of searching novel algorithmic applications of QWs by utilizing spin degrees of freedom. 
Furthermore, the nontrivial QW dynamics in combination with spin-mixing that we elucidate in this paper will offer a clue to understanding the equilibration or thermalization processes in spinfull systems.

\begin{acknowledgments}
This work was supported by Japan Society for the Promotion of Science KAKENHI Grant No. 25287104. 
\end{acknowledgments}

\appendix
\section{\label{app:EH}The Effective Hamiltonian}
Here we derive \Eqref{eq:FB} in an alternative way. We introduce the following effective Hamiltonian, which describes a strongly correlated boson pair in the bound state: 
\begin{align}
\hat{H}_{\rm eff}^B=J_U\sum_j(\hat{R}_j^{\dagger}\hat{R}_{j+1}+{\rm h.c})+(U+2J_U)\sum_j\hat{R}_j^{\dagger}\hat{R}_j,
\label{eq:EH}
\end{align}
where $J_U$ is the cotunneling amplitude, and $\hat{R}_j=(\hat{b}_j)^2/\sqrt{2}$ represents the annihilation operator of a boson pair at the $i$-th site. 
Note that in deriving \Eqref{eq:EH} we should retain the constant energy-shift that explicitly depends on the interaction strength. 
On the basis of this Hamiltonian, we obtain the dynamical evolution of the atom pair that is initially located at the origin,
\begin{align}
\Psi^B_U(R,r,t)&\simeq\delta_{r,0}e^{-iUt/\hbar}e^{-2iJ_Ut}i^{-R}{\cal J}_{R}(2J_Ut).
\end{align}
This is the wave function of single-particle continuous-time QWs. Then it is straightforward to derive
\begin{align}
X_B(t)&=\sum_{R,r}\hspace{0em}'\re\left[\Psi_{U_0-2U_2}^B\hspace{0em}^*(R,r,t)\Psi_{U_0+U_2}^B(R,r,t)\right],\nonumber\\
&\simeq{\cal J}_0(2\epsilon t)\cos[(3U_2-2\epsilon)t]. 
\end{align}

\bibliography{spin1QW}

\begin{thebibliography}{54}%
\makeatletter
\providecommand \@ifxundefined [1]{%
 \@ifx{#1\undefined}
}%
\providecommand \@ifnum [1]{%
 \ifnum #1\expandafter \@firstoftwo
 \else \expandafter \@secondoftwo
 \fi
}%
\providecommand \@ifx [1]{%
 \ifx #1\expandafter \@firstoftwo
 \else \expandafter \@secondoftwo
 \fi
}%
\providecommand \natexlab [1]{#1}%
\providecommand \enquote  [1]{``#1''}%
\providecommand \bibnamefont  [1]{#1}%
\providecommand \bibfnamefont [1]{#1}%
\providecommand \citenamefont [1]{#1}%
\providecommand \href@noop [0]{\@secondoftwo}%
\providecommand \href [0]{\begingroup \@sanitize@url \@href}%
\providecommand \@href[1]{\@@startlink{#1}\@@href}%
\providecommand \@@href[1]{\endgroup#1\@@endlink}%
\providecommand \@sanitize@url [0]{\catcode `\\12\catcode `\$12\catcode
  `\&12\catcode `\#12\catcode `\^12\catcode `\_12\catcode `\%12\relax}%
\providecommand \@@startlink[1]{}%
\providecommand \@@endlink[0]{}%
\providecommand \url  [0]{\begingroup\@sanitize@url \@url }%
\providecommand \@url [1]{\endgroup\@href {#1}{\urlprefix }}%
\providecommand \urlprefix  [0]{URL }%
\providecommand \Eprint [0]{\href }%
\providecommand \doibase [0]{http://dx.doi.org/}%
\providecommand \selectlanguage [0]{\@gobble}%
\providecommand \bibinfo  [0]{\@secondoftwo}%
\providecommand \bibfield  [0]{\@secondoftwo}%
\providecommand \translation [1]{[#1]}%
\providecommand \BibitemOpen [0]{}%
\providecommand \bibitemStop [0]{}%
\providecommand \bibitemNoStop [0]{.\EOS\space}%
\providecommand \EOS [0]{\spacefactor3000\relax}%
\providecommand \BibitemShut  [1]{\csname bibitem#1\endcsname}%
\let\auto@bib@innerbib\@empty
\bibitem [{\citenamefont {Motwani}\ and\ \citenamefont
  {Raghavan}(1995)}]{opac-b1089638}%
  \BibitemOpen
  \bibfield  {author} {\bibinfo {author} {\bibfnamefont {Rajeev}\ \bibnamefont
  {Motwani}}\ and\ \bibinfo {author} {\bibfnamefont {Prabhakar}\ \bibnamefont
  {Raghavan}},\ }\href {http://opac.inria.fr/record=b1089638} {\emph {\bibinfo
  {title} {Randomized algorithms}}}\ (\bibinfo  {publisher} {Cambridge
  University Press},\ \bibinfo {year} {1995})\BibitemShut {NoStop}%
\bibitem [{\citenamefont {Ambainis}(2003)}]{AMBAINIS:2003aa}%
  \BibitemOpen
  \bibfield  {author} {\bibinfo {author} {\bibfnamefont {Andris}\ \bibnamefont
  {Ambainis}},\ }\bibfield  {title} {\enquote {\bibinfo {title} {Quantum walks
  and their algorithmic applications},}\ }\bibfield  {booktitle} {\emph
  {\bibinfo {booktitle} {International Journal of Quantum Information}},\
  }\href {\doibase 10.1142/S0219749903000383} {\bibfield  {journal} {\bibinfo
  {journal} {International Journal of Quantum Information}\ }\textbf {\bibinfo
  {volume} {01}},\ \bibinfo {pages} {507--518} (\bibinfo {year}
  {2003})}\BibitemShut {NoStop}%
\bibitem [{\citenamefont {Kempe}(2003)}]{doi:10.1080/00107151031000110776}%
  \BibitemOpen
  \bibfield  {author} {\bibinfo {author} {\bibfnamefont {J.}~\bibnamefont
  {Kempe}},\ }\bibfield  {title} {\enquote {\bibinfo {title} {Quantum random
  walks: An introductory overview},}\ }\href {\doibase
  10.1080/00107151031000110776} {\bibfield  {journal} {\bibinfo  {journal}
  {Contemporary Physics}\ }\textbf {\bibinfo {volume} {44}},\ \bibinfo {pages}
  {307--327} (\bibinfo {year} {2003})}\BibitemShut {NoStop}%
\bibitem [{\citenamefont {Kendon}(2007)}]{Kendon:2007:DQW:1348911.1348916}%
  \BibitemOpen
  \bibfield  {author} {\bibinfo {author} {\bibfnamefont {Viv}\ \bibnamefont
  {Kendon}},\ }\bibfield  {title} {\enquote {\bibinfo {title} {Decoherence in
  quantum walks -- a review},}\ }\href {\doibase 10.1017/S0960129507006354}
  {\bibfield  {journal} {\bibinfo  {journal} {Mathematical. Structures in Comp.
  Sci.}\ }\textbf {\bibinfo {volume} {17}},\ \bibinfo {pages} {1169--1220}
  (\bibinfo {year} {2007})}\BibitemShut {NoStop}%
\bibitem [{\citenamefont
  {Venegas-Andraca}(2012)}]{Venegas-Andraca:2012:QWC:2386737.2386759}%
  \BibitemOpen
  \bibfield  {author} {\bibinfo {author} {\bibfnamefont {Salvador~El\'{\i}as}\
  \bibnamefont {Venegas-Andraca}},\ }\bibfield  {title} {\enquote {\bibinfo
  {title} {Quantum walks: A comprehensive review},}\ }\href {\doibase
  10.1007/s11128-012-0432-5} {\bibfield  {journal} {\bibinfo  {journal}
  {Quantum Information Processing}\ }\textbf {\bibinfo {volume} {11}},\
  \bibinfo {pages} {1015--1106} (\bibinfo {year} {2012})}\BibitemShut {NoStop}%
\bibitem [{\citenamefont {Shenvi}\ \emph {et~al.}(2003)\citenamefont {Shenvi},
  \citenamefont {Kempe},\ and\ \citenamefont {Whaley}}]{PhysRevA.67.052307}%
  \BibitemOpen
  \bibfield  {author} {\bibinfo {author} {\bibfnamefont {Neil}\ \bibnamefont
  {Shenvi}}, \bibinfo {author} {\bibfnamefont {Julia}\ \bibnamefont {Kempe}}, \
  and\ \bibinfo {author} {\bibfnamefont {K.~Birgitta}\ \bibnamefont {Whaley}},\
  }\bibfield  {title} {\enquote {\bibinfo {title} {Quantum random-walk search
  algorithm},}\ }\href {\doibase 10.1103/PhysRevA.67.052307} {\bibfield
  {journal} {\bibinfo  {journal} {Phys. Rev. A}\ }\textbf {\bibinfo {volume}
  {67}},\ \bibinfo {pages} {052307} (\bibinfo {year} {2003})}\BibitemShut
  {NoStop}%
\bibitem [{\citenamefont {Childs}\ and\ \citenamefont
  {Goldstone}(2004)}]{PhysRevA.70.022314}%
  \BibitemOpen
  \bibfield  {author} {\bibinfo {author} {\bibfnamefont {Andrew~M.}\
  \bibnamefont {Childs}}\ and\ \bibinfo {author} {\bibfnamefont {Jeffrey}\
  \bibnamefont {Goldstone}},\ }\bibfield  {title} {\enquote {\bibinfo {title}
  {Spatial search by quantum walk},}\ }\href {\doibase
  10.1103/PhysRevA.70.022314} {\bibfield  {journal} {\bibinfo  {journal} {Phys.
  Rev. A}\ }\textbf {\bibinfo {volume} {70}},\ \bibinfo {pages} {022314}
  (\bibinfo {year} {2004})}\BibitemShut {NoStop}%
\bibitem [{\citenamefont {Childs}(2009)}]{PhysRevLett.102.180501}%
  \BibitemOpen
  \bibfield  {author} {\bibinfo {author} {\bibfnamefont {Andrew~M.}\
  \bibnamefont {Childs}},\ }\bibfield  {title} {\enquote {\bibinfo {title}
  {Universal computation by quantum walk},}\ }\href {\doibase
  10.1103/PhysRevLett.102.180501} {\bibfield  {journal} {\bibinfo  {journal}
  {Phys. Rev. Lett.}\ }\textbf {\bibinfo {volume} {102}},\ \bibinfo {pages}
  {180501} (\bibinfo {year} {2009})}\BibitemShut {NoStop}%
\bibitem [{\citenamefont {Lovett}\ \emph {et~al.}(2010)\citenamefont {Lovett},
  \citenamefont {Cooper}, \citenamefont {Everitt}, \citenamefont {Trevers},\
  and\ \citenamefont {Kendon}}]{PhysRevA.81.042330}%
  \BibitemOpen
  \bibfield  {author} {\bibinfo {author} {\bibfnamefont {Neil~B.}\ \bibnamefont
  {Lovett}}, \bibinfo {author} {\bibfnamefont {Sally}\ \bibnamefont {Cooper}},
  \bibinfo {author} {\bibfnamefont {Matthew}\ \bibnamefont {Everitt}}, \bibinfo
  {author} {\bibfnamefont {Matthew}\ \bibnamefont {Trevers}}, \ and\ \bibinfo
  {author} {\bibfnamefont {Viv}\ \bibnamefont {Kendon}},\ }\bibfield  {title}
  {\enquote {\bibinfo {title} {Universal quantum computation using the
  discrete-time quantum walk},}\ }\href {\doibase 10.1103/PhysRevA.81.042330}
  {\bibfield  {journal} {\bibinfo  {journal} {Phys. Rev. A}\ }\textbf {\bibinfo
  {volume} {81}},\ \bibinfo {pages} {042330} (\bibinfo {year}
  {2010})}\BibitemShut {NoStop}%
\bibitem [{\citenamefont {Childs}\ \emph {et~al.}(2013)\citenamefont {Childs},
  \citenamefont {Gosset},\ and\ \citenamefont {Webb}}]{Childs15022013}%
  \BibitemOpen
  \bibfield  {author} {\bibinfo {author} {\bibfnamefont {Andrew~M.}\
  \bibnamefont {Childs}}, \bibinfo {author} {\bibfnamefont {David}\
  \bibnamefont {Gosset}}, \ and\ \bibinfo {author} {\bibfnamefont {Zak}\
  \bibnamefont {Webb}},\ }\bibfield  {title} {\enquote {\bibinfo {title}
  {Universal computation by multiparticle quantum walk},}\ }\href {\doibase
  10.1126/science.1229957} {\bibfield  {journal} {\bibinfo  {journal}
  {Science}\ }\textbf {\bibinfo {volume} {339}},\ \bibinfo {pages} {791--794}
  (\bibinfo {year} {2013})}\BibitemShut {NoStop}%
\bibitem [{\citenamefont {Aharonov}\ \emph {et~al.}(2001)\citenamefont
  {Aharonov}, \citenamefont {Ambainis}, \citenamefont {Kempe},\ and\
  \citenamefont {Vazirani}}]{Aharonov:2001:QWG:380752.380758}%
  \BibitemOpen
  \bibfield  {author} {\bibinfo {author} {\bibfnamefont {Dorit}\ \bibnamefont
  {Aharonov}}, \bibinfo {author} {\bibfnamefont {Andris}\ \bibnamefont
  {Ambainis}}, \bibinfo {author} {\bibfnamefont {Julia}\ \bibnamefont {Kempe}},
  \ and\ \bibinfo {author} {\bibfnamefont {Umesh}\ \bibnamefont {Vazirani}},\
  }\bibfield  {title} {\enquote {\bibinfo {title} {Quantum walks on graphs},}\
  }in\ \href {\doibase 10.1145/380752.380758} {\emph {\bibinfo {booktitle}
  {Proceedings of the Thirty-third Annual ACM Symposium on Theory of
  Computing}}},\ \bibinfo {series and number} {STOC '01}\ (\bibinfo
  {publisher} {ACM},\ \bibinfo {address} {New York, NY, USA},\ \bibinfo {year}
  {2001})\ pp.\ \bibinfo {pages} {50--59}\BibitemShut {NoStop}%
\bibitem [{\citenamefont {Ambainis}\ \emph {et~al.}(2001)\citenamefont
  {Ambainis}, \citenamefont {Bach}, \citenamefont {Nayak}, \citenamefont
  {Vishwanath},\ and\ \citenamefont
  {Watrous}}]{Ambainis:2001:OQW:380752.380757}%
  \BibitemOpen
  \bibfield  {author} {\bibinfo {author} {\bibfnamefont {Andris}\ \bibnamefont
  {Ambainis}}, \bibinfo {author} {\bibfnamefont {Eric}\ \bibnamefont {Bach}},
  \bibinfo {author} {\bibfnamefont {Ashwin}\ \bibnamefont {Nayak}}, \bibinfo
  {author} {\bibfnamefont {Ashvin}\ \bibnamefont {Vishwanath}}, \ and\ \bibinfo
  {author} {\bibfnamefont {John}\ \bibnamefont {Watrous}},\ }\bibfield  {title}
  {\enquote {\bibinfo {title} {One-dimensional quantum walks},}\ }in\ \href
  {\doibase 10.1145/380752.380757} {\emph {\bibinfo {booktitle} {Proceedings of
  the Thirty-third Annual ACM Symposium on Theory of Computing}}},\ \bibinfo
  {series and number} {STOC '01}\ (\bibinfo  {publisher} {ACM},\ \bibinfo
  {address} {New York, NY, USA},\ \bibinfo {year} {2001})\ pp.\ \bibinfo
  {pages} {37--49}\BibitemShut {NoStop}%
\bibitem [{\citenamefont {Farhi}\ and\ \citenamefont
  {Gutmann}(1998)}]{PhysRevA.58.915}%
  \BibitemOpen
  \bibfield  {author} {\bibinfo {author} {\bibfnamefont {E.}~\bibnamefont
  {Farhi}}\ and\ \bibinfo {author} {\bibfnamefont {S.}~\bibnamefont
  {Gutmann}},\ }\bibfield  {title} {\enquote {\bibinfo {title} {Quantum
  computation and decision trees},}\ }\href {\doibase 10.1103/PhysRevA.58.915}
  {\bibfield  {journal} {\bibinfo  {journal} {Phys. Rev. A}\ }\textbf {\bibinfo
  {volume} {58}},\ \bibinfo {pages} {915--928} (\bibinfo {year}
  {1998})}\BibitemShut {NoStop}%
\bibitem [{\citenamefont {Manouchehri}\ and\ \citenamefont
  {Wang}(2013)}]{Manouchehri:2013:PIQ:2566741}%
  \BibitemOpen
  \bibfield  {author} {\bibinfo {author} {\bibfnamefont {Kia}\ \bibnamefont
  {Manouchehri}}\ and\ \bibinfo {author} {\bibfnamefont {Jingbo}\ \bibnamefont
  {Wang}},\ }\href@noop {} {\emph {\bibinfo {title} {Physical Implementation of
  Quantum Walks}}}\ (\bibinfo  {publisher} {Springer Publishing Company,
  Incorporated},\ \bibinfo {year} {2013})\BibitemShut {NoStop}%
\bibitem [{\citenamefont {Bromberg}\ \emph {et~al.}(2009)\citenamefont
  {Bromberg}, \citenamefont {Lahini}, \citenamefont {Morandotti},\ and\
  \citenamefont {Silberberg}}]{PhysRevLett.102.253904}%
  \BibitemOpen
  \bibfield  {author} {\bibinfo {author} {\bibfnamefont {Yaron}\ \bibnamefont
  {Bromberg}}, \bibinfo {author} {\bibfnamefont {Yoav}\ \bibnamefont {Lahini}},
  \bibinfo {author} {\bibfnamefont {Roberto}\ \bibnamefont {Morandotti}}, \
  and\ \bibinfo {author} {\bibfnamefont {Yaron}\ \bibnamefont {Silberberg}},\
  }\bibfield  {title} {\enquote {\bibinfo {title} {Quantum and classical
  correlations in waveguide lattices},}\ }\href {\doibase
  10.1103/PhysRevLett.102.253904} {\bibfield  {journal} {\bibinfo  {journal}
  {Phys. Rev. Lett.}\ }\textbf {\bibinfo {volume} {102}},\ \bibinfo {pages}
  {253904} (\bibinfo {year} {2009})}\BibitemShut {NoStop}%
\bibitem [{\citenamefont {Peruzzo}\ \emph {et~al.}(2010)\citenamefont
  {Peruzzo}, \citenamefont {Lobino}, \citenamefont {Matthews}, \citenamefont
  {Matsuda}, \citenamefont {Politi}, \citenamefont {Poulios}, \citenamefont
  {Zhou}, \citenamefont {Lahini}, \citenamefont {Ismail}, \citenamefont
  {W{\"o}rhoff}, \citenamefont {Bromberg}, \citenamefont {Silberberg},
  \citenamefont {Thompson},\ and\ \citenamefont {OBrien}}]{Peruzzo1500}%
  \BibitemOpen
  \bibfield  {author} {\bibinfo {author} {\bibfnamefont {Alberto}\ \bibnamefont
  {Peruzzo}}, \bibinfo {author} {\bibfnamefont {Mirko}\ \bibnamefont {Lobino}},
  \bibinfo {author} {\bibfnamefont {Jonathan C.~F.}\ \bibnamefont {Matthews}},
  \bibinfo {author} {\bibfnamefont {Nobuyuki}\ \bibnamefont {Matsuda}},
  \bibinfo {author} {\bibfnamefont {Alberto}\ \bibnamefont {Politi}}, \bibinfo
  {author} {\bibfnamefont {Konstantinos}\ \bibnamefont {Poulios}}, \bibinfo
  {author} {\bibfnamefont {Xiao-Qi}\ \bibnamefont {Zhou}}, \bibinfo {author}
  {\bibfnamefont {Yoav}\ \bibnamefont {Lahini}}, \bibinfo {author}
  {\bibfnamefont {Nur}\ \bibnamefont {Ismail}}, \bibinfo {author}
  {\bibfnamefont {Kerstin}\ \bibnamefont {W{\"o}rhoff}}, \bibinfo {author}
  {\bibfnamefont {Yaron}\ \bibnamefont {Bromberg}}, \bibinfo {author}
  {\bibfnamefont {Yaron}\ \bibnamefont {Silberberg}}, \bibinfo {author}
  {\bibfnamefont {Mark~G.}\ \bibnamefont {Thompson}}, \ and\ \bibinfo {author}
  {\bibfnamefont {Jeremy~L.}\ \bibnamefont {OBrien}},\ }\bibfield  {title}
  {\enquote {\bibinfo {title} {Quantum walks of correlated photons},}\ }\href
  {\doibase 10.1126/science.1193515} {\bibfield  {journal} {\bibinfo  {journal}
  {Science}\ }\textbf {\bibinfo {volume} {329}},\ \bibinfo {pages} {1500--1503}
  (\bibinfo {year} {2010})}\BibitemShut {NoStop}%
\bibitem [{\citenamefont {Lahini}\ \emph {et~al.}(2012)\citenamefont {Lahini},
  \citenamefont {Verbin}, \citenamefont {Huber}, \citenamefont {Bromberg},
  \citenamefont {Pugatch},\ and\ \citenamefont
  {Silberberg}}]{PhysRevA.86.011603}%
  \BibitemOpen
  \bibfield  {author} {\bibinfo {author} {\bibfnamefont {Yoav}\ \bibnamefont
  {Lahini}}, \bibinfo {author} {\bibfnamefont {Mor}\ \bibnamefont {Verbin}},
  \bibinfo {author} {\bibfnamefont {Sebastian~D.}\ \bibnamefont {Huber}},
  \bibinfo {author} {\bibfnamefont {Yaron}\ \bibnamefont {Bromberg}}, \bibinfo
  {author} {\bibfnamefont {Rami}\ \bibnamefont {Pugatch}}, \ and\ \bibinfo
  {author} {\bibfnamefont {Yaron}\ \bibnamefont {Silberberg}},\ }\bibfield
  {title} {\enquote {\bibinfo {title} {Quantum walk of two interacting
  bosons},}\ }\href {\doibase 10.1103/PhysRevA.86.011603} {\bibfield  {journal}
  {\bibinfo  {journal} {Phys. Rev. A}\ }\textbf {\bibinfo {volume} {86}},\
  \bibinfo {pages} {011603} (\bibinfo {year} {2012})}\BibitemShut {NoStop}%
\bibitem [{\citenamefont {Preiss}\ \emph {et~al.}(2015)\citenamefont {Preiss},
  \citenamefont {Ma}, \citenamefont {Tai}, \citenamefont {Lukin}, \citenamefont
  {Rispoli}, \citenamefont {Zupancic}, \citenamefont {Lahini}, \citenamefont
  {Islam},\ and\ \citenamefont {Greiner}}]{Preiss13032015}%
  \BibitemOpen
  \bibfield  {author} {\bibinfo {author} {\bibfnamefont {Philipp~M.}\
  \bibnamefont {Preiss}}, \bibinfo {author} {\bibfnamefont {Ruichao}\
  \bibnamefont {Ma}}, \bibinfo {author} {\bibfnamefont {M.~Eric}\ \bibnamefont
  {Tai}}, \bibinfo {author} {\bibfnamefont {Alexander}\ \bibnamefont {Lukin}},
  \bibinfo {author} {\bibfnamefont {Matthew}\ \bibnamefont {Rispoli}}, \bibinfo
  {author} {\bibfnamefont {Philip}\ \bibnamefont {Zupancic}}, \bibinfo {author}
  {\bibfnamefont {Yoav}\ \bibnamefont {Lahini}}, \bibinfo {author}
  {\bibfnamefont {Rajibul}\ \bibnamefont {Islam}}, \ and\ \bibinfo {author}
  {\bibfnamefont {Markus}\ \bibnamefont {Greiner}},\ }\bibfield  {title}
  {\enquote {\bibinfo {title} {Strongly correlated quantum walks in optical
  lattices},}\ }\href {\doibase 10.1126/science.1260364} {\bibfield  {journal}
  {\bibinfo  {journal} {Science}\ }\textbf {\bibinfo {volume} {347}},\ \bibinfo
  {pages} {1229--1233} (\bibinfo {year} {2015})}\BibitemShut {NoStop}%
\bibitem [{\citenamefont {Bloch}\ \emph {et~al.}(2012)\citenamefont {Bloch},
  \citenamefont {Dalibard},\ and\ \citenamefont {Nascimbene}}]{Bloch:2012aa}%
  \BibitemOpen
  \bibfield  {author} {\bibinfo {author} {\bibfnamefont {Immanuel}\
  \bibnamefont {Bloch}}, \bibinfo {author} {\bibfnamefont {Jean}\ \bibnamefont
  {Dalibard}}, \ and\ \bibinfo {author} {\bibfnamefont {Sylvain}\ \bibnamefont
  {Nascimbene}},\ }\bibfield  {title} {\enquote {\bibinfo {title} {Quantum
  simulations with ultracold quantum gases},}\ }\href
  {http://dx.doi.org/10.1038/nphys2259} {\bibfield  {journal} {\bibinfo
  {journal} {Nat Phys}\ }\textbf {\bibinfo {volume} {8}},\ \bibinfo {pages}
  {267--276} (\bibinfo {year} {2012})}\BibitemShut {NoStop}%
\bibitem [{\citenamefont {Georgescu}\ \emph {et~al.}(2014)\citenamefont
  {Georgescu}, \citenamefont {Ashhab},\ and\ \citenamefont
  {Nori}}]{RevModPhys.86.153}%
  \BibitemOpen
  \bibfield  {author} {\bibinfo {author} {\bibfnamefont {I.~M.}\ \bibnamefont
  {Georgescu}}, \bibinfo {author} {\bibfnamefont {S.}~\bibnamefont {Ashhab}}, \
  and\ \bibinfo {author} {\bibfnamefont {Franco}\ \bibnamefont {Nori}},\
  }\bibfield  {title} {\enquote {\bibinfo {title} {Quantum simulation},}\
  }\href {\doibase 10.1103/RevModPhys.86.153} {\bibfield  {journal} {\bibinfo
  {journal} {Rev. Mod. Phys.}\ }\textbf {\bibinfo {volume} {86}},\ \bibinfo
  {pages} {153--185} (\bibinfo {year} {2014})}\BibitemShut {NoStop}%
\bibitem [{\citenamefont {Sherson}\ \emph {et~al.}(2010)\citenamefont
  {Sherson}, \citenamefont {Weitenberg}, \citenamefont {Endres}, \citenamefont
  {Cheneau}, \citenamefont {Bloch},\ and\ \citenamefont
  {Kuhr}}]{Sherson:2010aa}%
  \BibitemOpen
  \bibfield  {author} {\bibinfo {author} {\bibfnamefont {Jacob~F.}\
  \bibnamefont {Sherson}}, \bibinfo {author} {\bibfnamefont {Christof}\
  \bibnamefont {Weitenberg}}, \bibinfo {author} {\bibfnamefont {Manuel}\
  \bibnamefont {Endres}}, \bibinfo {author} {\bibfnamefont {Marc}\ \bibnamefont
  {Cheneau}}, \bibinfo {author} {\bibfnamefont {Immanuel}\ \bibnamefont
  {Bloch}}, \ and\ \bibinfo {author} {\bibfnamefont {Stefan}\ \bibnamefont
  {Kuhr}},\ }\bibfield  {title} {\enquote {\bibinfo {title}
  {Single-atom-resolved fluorescence imaging of an atomic {M}ott insulator},}\
  }\href {http://dx.doi.org/10.1038/nature09378} {\bibfield  {journal}
  {\bibinfo  {journal} {Nature}\ }\textbf {\bibinfo {volume} {467}},\ \bibinfo
  {pages} {68--72} (\bibinfo {year} {2010})}\BibitemShut {NoStop}%
\bibitem [{\citenamefont {Weitenberg}\ \emph {et~al.}(2011)\citenamefont
  {Weitenberg}, \citenamefont {Endres}, \citenamefont {Sherson}, \citenamefont
  {Cheneau}, \citenamefont {Schausz}, \citenamefont {Fukuhara}, \citenamefont
  {Bloch},\ and\ \citenamefont {Kuhr}}]{Weitenberg:2011aa}%
  \BibitemOpen
  \bibfield  {author} {\bibinfo {author} {\bibfnamefont {Christof}\
  \bibnamefont {Weitenberg}}, \bibinfo {author} {\bibfnamefont {Manuel}\
  \bibnamefont {Endres}}, \bibinfo {author} {\bibfnamefont {Jacob~F.}\
  \bibnamefont {Sherson}}, \bibinfo {author} {\bibfnamefont {Marc}\
  \bibnamefont {Cheneau}}, \bibinfo {author} {\bibfnamefont {Peter}\
  \bibnamefont {Schausz}}, \bibinfo {author} {\bibfnamefont {Takeshi}\
  \bibnamefont {Fukuhara}}, \bibinfo {author} {\bibfnamefont {Immanuel}\
  \bibnamefont {Bloch}}, \ and\ \bibinfo {author} {\bibfnamefont {Stefan}\
  \bibnamefont {Kuhr}},\ }\bibfield  {title} {\enquote {\bibinfo {title}
  {Single-spin addressing in an atomic {M}ott insulator},}\ }\href
  {http://dx.doi.org/10.1038/nature09827} {\bibfield  {journal} {\bibinfo
  {journal} {Nature}\ }\textbf {\bibinfo {volume} {471}},\ \bibinfo {pages}
  {319--324} (\bibinfo {year} {2011})}\BibitemShut {NoStop}%
\bibitem [{\citenamefont {Fukuhara}\ \emph {et~al.}(2013)\citenamefont
  {Fukuhara}, \citenamefont {Schausz}, \citenamefont {Endres}, \citenamefont
  {Hild}, \citenamefont {Cheneau}, \citenamefont {Bloch},\ and\ \citenamefont
  {Gross}}]{Fukuhara:2013ab}%
  \BibitemOpen
  \bibfield  {author} {\bibinfo {author} {\bibfnamefont {Takeshi}\ \bibnamefont
  {Fukuhara}}, \bibinfo {author} {\bibfnamefont {Peter}\ \bibnamefont
  {Schausz}}, \bibinfo {author} {\bibfnamefont {Manuel}\ \bibnamefont
  {Endres}}, \bibinfo {author} {\bibfnamefont {Sebastian}\ \bibnamefont
  {Hild}}, \bibinfo {author} {\bibfnamefont {Marc}\ \bibnamefont {Cheneau}},
  \bibinfo {author} {\bibfnamefont {Immanuel}\ \bibnamefont {Bloch}}, \ and\
  \bibinfo {author} {\bibfnamefont {Christian}\ \bibnamefont {Gross}},\
  }\bibfield  {title} {\enquote {\bibinfo {title} {Microscopic observation of
  magnon bound states and their dynamics},}\ }\href
  {http://dx.doi.org/10.1038/nature12541} {\bibfield  {journal} {\bibinfo
  {journal} {Nature}\ }\textbf {\bibinfo {volume} {502}},\ \bibinfo {pages}
  {76--79} (\bibinfo {year} {2013})}\BibitemShut {NoStop}%
\bibitem [{\citenamefont {Stamper-Kurn}\ \emph {et~al.}(1998)\citenamefont
  {Stamper-Kurn}, \citenamefont {Andrews}, \citenamefont {Chikkatur},
  \citenamefont {Inouye}, \citenamefont {Miesner}, \citenamefont {Stenger},\
  and\ \citenamefont {Ketterle}}]{PhysRevLett.80.2027}%
  \BibitemOpen
  \bibfield  {author} {\bibinfo {author} {\bibfnamefont {D.~M.}\ \bibnamefont
  {Stamper-Kurn}}, \bibinfo {author} {\bibfnamefont {M.~R.}\ \bibnamefont
  {Andrews}}, \bibinfo {author} {\bibfnamefont {A.~P.}\ \bibnamefont
  {Chikkatur}}, \bibinfo {author} {\bibfnamefont {S.}~\bibnamefont {Inouye}},
  \bibinfo {author} {\bibfnamefont {H.-J.}\ \bibnamefont {Miesner}}, \bibinfo
  {author} {\bibfnamefont {J.}~\bibnamefont {Stenger}}, \ and\ \bibinfo
  {author} {\bibfnamefont {W.}~\bibnamefont {Ketterle}},\ }\bibfield  {title}
  {\enquote {\bibinfo {title} {Optical confinement of a {B}ose-{E}instein
  condensate},}\ }\href {\doibase 10.1103/PhysRevLett.80.2027} {\bibfield
  {journal} {\bibinfo  {journal} {Phys. Rev. Lett.}\ }\textbf {\bibinfo
  {volume} {80}},\ \bibinfo {pages} {2027--2030} (\bibinfo {year}
  {1998})}\BibitemShut {NoStop}%
\bibitem [{\citenamefont {Ho}(1998)}]{PhysRevLett.81.742}%
  \BibitemOpen
  \bibfield  {author} {\bibinfo {author} {\bibfnamefont {Tin-Lun}\ \bibnamefont
  {Ho}},\ }\bibfield  {title} {\enquote {\bibinfo {title} {Spinor {B}ose
  condensates in optical traps},}\ }\href {\doibase 10.1103/PhysRevLett.81.742}
  {\bibfield  {journal} {\bibinfo  {journal} {Phys. Rev. Lett.}\ }\textbf
  {\bibinfo {volume} {81}},\ \bibinfo {pages} {742--745} (\bibinfo {year}
  {1998})}\BibitemShut {NoStop}%
\bibitem [{\citenamefont {Barrett}\ \emph {et~al.}(2001)\citenamefont
  {Barrett}, \citenamefont {Sauer},\ and\ \citenamefont
  {Chapman}}]{PhysRevLett.87.010404}%
  \BibitemOpen
  \bibfield  {author} {\bibinfo {author} {\bibfnamefont {M.~D.}\ \bibnamefont
  {Barrett}}, \bibinfo {author} {\bibfnamefont {J.~A.}\ \bibnamefont {Sauer}},
  \ and\ \bibinfo {author} {\bibfnamefont {M.~S.}\ \bibnamefont {Chapman}},\
  }\bibfield  {title} {\enquote {\bibinfo {title} {All-optical formation of an
  atomic {B}ose-{E}instein condensate},}\ }\href {\doibase
  10.1103/PhysRevLett.87.010404} {\bibfield  {journal} {\bibinfo  {journal}
  {Phys. Rev. Lett.}\ }\textbf {\bibinfo {volume} {87}},\ \bibinfo {pages}
  {010404} (\bibinfo {year} {2001})}\BibitemShut {NoStop}%
\bibitem [{\citenamefont {Kawaguchi}\ and\ \citenamefont
  {Ueda}(2012)}]{Kawaguchi2012253}%
  \BibitemOpen
  \bibfield  {author} {\bibinfo {author} {\bibfnamefont {Yuki}\ \bibnamefont
  {Kawaguchi}}\ and\ \bibinfo {author} {\bibfnamefont {Masahito}\ \bibnamefont
  {Ueda}},\ }\bibfield  {title} {\enquote {\bibinfo {title} {Spinor
  {B}ose-{E}instein condensates},}\ }\href {\doibase
  http://dx.doi.org/10.1016/j.physrep.2012.07.005} {\bibfield  {journal}
  {\bibinfo  {journal} {Physics Reports}\ }\textbf {\bibinfo {volume} {520}},\
  \bibinfo {pages} {253 -- 381} (\bibinfo {year} {2012})}\BibitemShut {NoStop}%
\bibitem [{\citenamefont {Stamper-Kurn}\ and\ \citenamefont
  {Ueda}(2013)}]{RevModPhys.85.1191}%
  \BibitemOpen
  \bibfield  {author} {\bibinfo {author} {\bibfnamefont {Dan~M.}\ \bibnamefont
  {Stamper-Kurn}}\ and\ \bibinfo {author} {\bibfnamefont {Masahito}\
  \bibnamefont {Ueda}},\ }\bibfield  {title} {\enquote {\bibinfo {title}
  {Spinor {B}ose gases: Symmetries, magnetism, and quantum dynamics},}\ }\href
  {\doibase 10.1103/RevModPhys.85.1191} {\bibfield  {journal} {\bibinfo
  {journal} {Rev. Mod. Phys.}\ }\textbf {\bibinfo {volume} {85}},\ \bibinfo
  {pages} {1191--1244} (\bibinfo {year} {2013})}\BibitemShut {NoStop}%
\bibitem [{\citenamefont {Demler}\ and\ \citenamefont
  {Zhou}(2002)}]{PhysRevLett.88.163001}%
  \BibitemOpen
  \bibfield  {author} {\bibinfo {author} {\bibfnamefont {Eugene}\ \bibnamefont
  {Demler}}\ and\ \bibinfo {author} {\bibfnamefont {Fei}\ \bibnamefont
  {Zhou}},\ }\bibfield  {title} {\enquote {\bibinfo {title} {Spinor bosonic
  atoms in optical lattices: Symmetry breaking and fractionalization},}\ }\href
  {\doibase 10.1103/PhysRevLett.88.163001} {\bibfield  {journal} {\bibinfo
  {journal} {Phys. Rev. Lett.}\ }\textbf {\bibinfo {volume} {88}},\ \bibinfo
  {pages} {163001} (\bibinfo {year} {2002})}\BibitemShut {NoStop}%
\bibitem [{\citenamefont {Imambekov}\ \emph {et~al.}(2003)\citenamefont
  {Imambekov}, \citenamefont {Lukin},\ and\ \citenamefont
  {Demler}}]{PhysRevA.68.063602}%
  \BibitemOpen
  \bibfield  {author} {\bibinfo {author} {\bibfnamefont {Adilet}\ \bibnamefont
  {Imambekov}}, \bibinfo {author} {\bibfnamefont {Mikhail}\ \bibnamefont
  {Lukin}}, \ and\ \bibinfo {author} {\bibfnamefont {Eugene}\ \bibnamefont
  {Demler}},\ }\bibfield  {title} {\enquote {\bibinfo {title} {Spin-exchange
  interactions of spin-one bosons in optical lattices: Singlet, nematic, and
  dimerized phases},}\ }\href {\doibase 10.1103/PhysRevA.68.063602} {\bibfield
  {journal} {\bibinfo  {journal} {Phys. Rev. A}\ }\textbf {\bibinfo {volume}
  {68}},\ \bibinfo {pages} {063602} (\bibinfo {year} {2003})}\BibitemShut
  {NoStop}%
\bibitem [{\citenamefont {Snoek}\ and\ \citenamefont
  {Zhou}(2004)}]{PhysRevB.69.094410}%
  \BibitemOpen
  \bibfield  {author} {\bibinfo {author} {\bibfnamefont {Michiel}\ \bibnamefont
  {Snoek}}\ and\ \bibinfo {author} {\bibfnamefont {Fei}\ \bibnamefont {Zhou}},\
  }\bibfield  {title} {\enquote {\bibinfo {title} {Microscopic wave functions
  of spin-singlet and nematic {M}ott states of spin-one bosons in
  high-dimensional bipartite lattices},}\ }\href {\doibase
  10.1103/PhysRevB.69.094410} {\bibfield  {journal} {\bibinfo  {journal} {Phys.
  Rev. B}\ }\textbf {\bibinfo {volume} {69}},\ \bibinfo {pages} {094410}
  (\bibinfo {year} {2004})}\BibitemShut {NoStop}%
\bibitem [{\citenamefont {Tsuchiya}\ \emph {et~al.}(2004)\citenamefont
  {Tsuchiya}, \citenamefont {Kurihara},\ and\ \citenamefont
  {Kimura}}]{PhysRevA.70.043628}%
  \BibitemOpen
  \bibfield  {author} {\bibinfo {author} {\bibfnamefont {Shunji}\ \bibnamefont
  {Tsuchiya}}, \bibinfo {author} {\bibfnamefont {Susumu}\ \bibnamefont
  {Kurihara}}, \ and\ \bibinfo {author} {\bibfnamefont {Takashi}\ \bibnamefont
  {Kimura}},\ }\bibfield  {title} {\enquote {\bibinfo {title}
  {Superfluid-{M}ott insulator transition of spin-1 bosons in an optical
  lattice},}\ }\href {\doibase 10.1103/PhysRevA.70.043628} {\bibfield
  {journal} {\bibinfo  {journal} {Phys. Rev. A}\ }\textbf {\bibinfo {volume}
  {70}},\ \bibinfo {pages} {043628} (\bibinfo {year} {2004})}\BibitemShut
  {NoStop}%
\bibitem [{\citenamefont {Krutitsky}\ and\ \citenamefont
  {Graham}(2004)}]{PhysRevA.70.063610}%
  \BibitemOpen
  \bibfield  {author} {\bibinfo {author} {\bibfnamefont {K.~V.}\ \bibnamefont
  {Krutitsky}}\ and\ \bibinfo {author} {\bibfnamefont {R.}~\bibnamefont
  {Graham}},\ }\bibfield  {title} {\enquote {\bibinfo {title} {Spin-1 bosons
  with coupled ground states in optical lattices},}\ }\href {\doibase
  10.1103/PhysRevA.70.063610} {\bibfield  {journal} {\bibinfo  {journal} {Phys.
  Rev. A}\ }\textbf {\bibinfo {volume} {70}},\ \bibinfo {pages} {063610}
  (\bibinfo {year} {2004})}\BibitemShut {NoStop}%
\bibitem [{\citenamefont {Kimura}\ \emph {et~al.}(2005)\citenamefont {Kimura},
  \citenamefont {Tsuchiya},\ and\ \citenamefont
  {Kurihara}}]{PhysRevLett.94.110403}%
  \BibitemOpen
  \bibfield  {author} {\bibinfo {author} {\bibfnamefont {Takashi}\ \bibnamefont
  {Kimura}}, \bibinfo {author} {\bibfnamefont {Shunji}\ \bibnamefont
  {Tsuchiya}}, \ and\ \bibinfo {author} {\bibfnamefont {Susumu}\ \bibnamefont
  {Kurihara}},\ }\bibfield  {title} {\enquote {\bibinfo {title} {Possibility of
  a first-order superfluid-{M}ott-insulator transition of spinor bosons in an
  optical lattice},}\ }\href {\doibase 10.1103/PhysRevLett.94.110403}
  {\bibfield  {journal} {\bibinfo  {journal} {Phys. Rev. Lett.}\ }\textbf
  {\bibinfo {volume} {94}},\ \bibinfo {pages} {110403} (\bibinfo {year}
  {2005})}\BibitemShut {NoStop}%
\bibitem [{\citenamefont {Rizzi}\ \emph {et~al.}(2005)\citenamefont {Rizzi},
  \citenamefont {Rossini}, \citenamefont {De~Chiara}, \citenamefont
  {Montangero},\ and\ \citenamefont {Fazio}}]{PhysRevLett.95.240404}%
  \BibitemOpen
  \bibfield  {author} {\bibinfo {author} {\bibfnamefont {Matteo}\ \bibnamefont
  {Rizzi}}, \bibinfo {author} {\bibfnamefont {Davide}\ \bibnamefont {Rossini}},
  \bibinfo {author} {\bibfnamefont {Gabriele}\ \bibnamefont {De~Chiara}},
  \bibinfo {author} {\bibfnamefont {Simone}\ \bibnamefont {Montangero}}, \ and\
  \bibinfo {author} {\bibfnamefont {Rosario}\ \bibnamefont {Fazio}},\
  }\bibfield  {title} {\enquote {\bibinfo {title} {Phase diagram of spin-1
  bosons on one-dimensional lattices},}\ }\href {\doibase
  10.1103/PhysRevLett.95.240404} {\bibfield  {journal} {\bibinfo  {journal}
  {Phys. Rev. Lett.}\ }\textbf {\bibinfo {volume} {95}},\ \bibinfo {pages}
  {240404} (\bibinfo {year} {2005})}\BibitemShut {NoStop}%
\bibitem [{\citenamefont {Apaja}\ and\ \citenamefont
  {Sylju\aa{}sen}(2006)}]{PhysRevA.74.035601}%
  \BibitemOpen
  \bibfield  {author} {\bibinfo {author} {\bibfnamefont {Vesa}\ \bibnamefont
  {Apaja}}\ and\ \bibinfo {author} {\bibfnamefont {Olav~F.}\ \bibnamefont
  {Sylju\aa{}sen}},\ }\bibfield  {title} {\enquote {\bibinfo {title} {Dimerized
  ground state in the one-dimensional spin-1 boson hubbard model},}\ }\href
  {\doibase 10.1103/PhysRevA.74.035601} {\bibfield  {journal} {\bibinfo
  {journal} {Phys. Rev. A}\ }\textbf {\bibinfo {volume} {74}},\ \bibinfo
  {pages} {035601} (\bibinfo {year} {2006})}\BibitemShut {NoStop}%
\bibitem [{\citenamefont {Yamashita}\ and\ \citenamefont
  {Jack}(2007)}]{PhysRevA.76.023606}%
  \BibitemOpen
  \bibfield  {author} {\bibinfo {author} {\bibfnamefont {Makoto}\ \bibnamefont
  {Yamashita}}\ and\ \bibinfo {author} {\bibfnamefont {Michael~W.}\
  \bibnamefont {Jack}},\ }\bibfield  {title} {\enquote {\bibinfo {title} {Spin
  structures of spin-1 bosonic atoms trapped in an optical lattice with
  harmonic confinement},}\ }\href {\doibase 10.1103/PhysRevA.76.023606}
  {\bibfield  {journal} {\bibinfo  {journal} {Phys. Rev. A}\ }\textbf {\bibinfo
  {volume} {76}},\ \bibinfo {pages} {023606} (\bibinfo {year}
  {2007})}\BibitemShut {NoStop}%
\bibitem [{\citenamefont {Toga}\ \emph {et~al.}(2012)\citenamefont {Toga},
  \citenamefont {Tsuchiura}, \citenamefont {Yamashita}, \citenamefont {Inaba},\
  and\ \citenamefont {Yokoyama}}]{Toga:2012aa}%
  \BibitemOpen
  \bibfield  {author} {\bibinfo {author} {\bibfnamefont {Yuta}\ \bibnamefont
  {Toga}}, \bibinfo {author} {\bibfnamefont {Hiroki}\ \bibnamefont
  {Tsuchiura}}, \bibinfo {author} {\bibfnamefont {Makoto}\ \bibnamefont
  {Yamashita}}, \bibinfo {author} {\bibfnamefont {Kensuke}\ \bibnamefont
  {Inaba}}, \ and\ \bibinfo {author} {\bibfnamefont {Hisatoshi}\ \bibnamefont
  {Yokoyama}},\ }\bibfield  {title} {\enquote {\bibinfo {title} {Mott
  transition and spin structures of spin-1 bosons in two-dimensional optical
  lattice at unit filling},}\ }\href {\doibase 10.1143/JPSJ.81.063001}
  {\bibfield  {journal} {\bibinfo  {journal} {Journal of the Physical Society
  of Japan}\ }\textbf {\bibinfo {volume} {81}},\ \bibinfo {pages} {063001}
  (\bibinfo {year} {2012})}\BibitemShut {NoStop}%
\bibitem [{\citenamefont {de~Forges~de Parny}\ \emph
  {et~al.}(2013)\citenamefont {de~Forges~de Parny}, \citenamefont {H\'ebert},
  \citenamefont {Rousseau},\ and\ \citenamefont
  {Batrouni}}]{PhysRevB.88.104509}%
  \BibitemOpen
  \bibfield  {author} {\bibinfo {author} {\bibfnamefont {L.}~\bibnamefont
  {de~Forges~de Parny}}, \bibinfo {author} {\bibfnamefont {F.}~\bibnamefont
  {H\'ebert}}, \bibinfo {author} {\bibfnamefont {V.~G.}\ \bibnamefont
  {Rousseau}}, \ and\ \bibinfo {author} {\bibfnamefont {G.~G.}\ \bibnamefont
  {Batrouni}},\ }\bibfield  {title} {\enquote {\bibinfo {title} {Interacting
  spin-1 bosons in a two-dimensional optical lattice},}\ }\href {\doibase
  10.1103/PhysRevB.88.104509} {\bibfield  {journal} {\bibinfo  {journal} {Phys.
  Rev. B}\ }\textbf {\bibinfo {volume} {88}},\ \bibinfo {pages} {104509}
  (\bibinfo {year} {2013})}\BibitemShut {NoStop}%
\bibitem [{\citenamefont {Law}\ \emph {et~al.}(1998)\citenamefont {Law},
  \citenamefont {Pu},\ and\ \citenamefont {Bigelow}}]{PhysRevLett.81.5257}%
  \BibitemOpen
  \bibfield  {author} {\bibinfo {author} {\bibfnamefont {C.~K.}\ \bibnamefont
  {Law}}, \bibinfo {author} {\bibfnamefont {H.}~\bibnamefont {Pu}}, \ and\
  \bibinfo {author} {\bibfnamefont {N.~P.}\ \bibnamefont {Bigelow}},\
  }\bibfield  {title} {\enquote {\bibinfo {title} {Quantum spins mixing in
  spinor {B}ose-{E}instein condensates},}\ }\href {\doibase
  10.1103/PhysRevLett.81.5257} {\bibfield  {journal} {\bibinfo  {journal}
  {Phys. Rev. Lett.}\ }\textbf {\bibinfo {volume} {81}},\ \bibinfo {pages}
  {5257--5261} (\bibinfo {year} {1998})}\BibitemShut {NoStop}%
\bibitem [{\citenamefont {Stenger}\ \emph {et~al.}(1998)\citenamefont
  {Stenger}, \citenamefont {Inouye}, \citenamefont {Stamper-Kurn},
  \citenamefont {Miesner}, \citenamefont {Chikkatur},\ and\ \citenamefont
  {Ketterle}}]{Stenger:1998aa}%
  \BibitemOpen
  \bibfield  {author} {\bibinfo {author} {\bibfnamefont {J.}~\bibnamefont
  {Stenger}}, \bibinfo {author} {\bibfnamefont {S.}~\bibnamefont {Inouye}},
  \bibinfo {author} {\bibfnamefont {D.~M.}\ \bibnamefont {Stamper-Kurn}},
  \bibinfo {author} {\bibfnamefont {H.~J.}\ \bibnamefont {Miesner}}, \bibinfo
  {author} {\bibfnamefont {A.~P.}\ \bibnamefont {Chikkatur}}, \ and\ \bibinfo
  {author} {\bibfnamefont {W.}~\bibnamefont {Ketterle}},\ }\bibfield  {title}
  {\enquote {\bibinfo {title} {Spin domains in ground-state {B}ose-{E}instein
  condensates},}\ }\href {http://dx.doi.org/10.1038/24567} {\bibfield
  {journal} {\bibinfo  {journal} {Nature}\ }\textbf {\bibinfo {volume} {396}},\
  \bibinfo {pages} {345--348} (\bibinfo {year} {1998})}\BibitemShut {NoStop}%
\bibitem [{\citenamefont {Chang}\ \emph {et~al.}(2005)\citenamefont {Chang},
  \citenamefont {Qin}, \citenamefont {Zhang}, \citenamefont {You},\ and\
  \citenamefont {Chapman}}]{Chang:2005aa}%
  \BibitemOpen
  \bibfield  {author} {\bibinfo {author} {\bibfnamefont {Ming-Shien}\
  \bibnamefont {Chang}}, \bibinfo {author} {\bibfnamefont {Qishu}\ \bibnamefont
  {Qin}}, \bibinfo {author} {\bibfnamefont {Wenxian}\ \bibnamefont {Zhang}},
  \bibinfo {author} {\bibfnamefont {Li}~\bibnamefont {You}}, \ and\ \bibinfo
  {author} {\bibfnamefont {Michael~S.}\ \bibnamefont {Chapman}},\ }\bibfield
  {title} {\enquote {\bibinfo {title} {Coherent spinor dynamics in a spin-1
  {B}ose condensate},}\ }\href {http://dx.doi.org/10.1038/nphys153} {\bibfield
  {journal} {\bibinfo  {journal} {Nat Phys}\ }\textbf {\bibinfo {volume} {1}},\
  \bibinfo {pages} {111--116} (\bibinfo {year} {2005})}\BibitemShut {NoStop}%
\bibitem [{\citenamefont {Kronj\"ager}\ \emph {et~al.}(2005)\citenamefont
  {Kronj\"ager}, \citenamefont {Becker}, \citenamefont {Brinkmann},
  \citenamefont {Walser}, \citenamefont {Navez}, \citenamefont {Bongs},\ and\
  \citenamefont {Sengstock}}]{PhysRevA.72.063619}%
  \BibitemOpen
  \bibfield  {author} {\bibinfo {author} {\bibfnamefont {J.}~\bibnamefont
  {Kronj\"ager}}, \bibinfo {author} {\bibfnamefont {C.}~\bibnamefont {Becker}},
  \bibinfo {author} {\bibfnamefont {M.}~\bibnamefont {Brinkmann}}, \bibinfo
  {author} {\bibfnamefont {R.}~\bibnamefont {Walser}}, \bibinfo {author}
  {\bibfnamefont {P.}~\bibnamefont {Navez}}, \bibinfo {author} {\bibfnamefont
  {K.}~\bibnamefont {Bongs}}, \ and\ \bibinfo {author} {\bibfnamefont
  {K.}~\bibnamefont {Sengstock}},\ }\bibfield  {title} {\enquote {\bibinfo
  {title} {Evolution of a spinor condensate: Coherent dynamics, dephasing, and
  revivals},}\ }\href {\doibase 10.1103/PhysRevA.72.063619} {\bibfield
  {journal} {\bibinfo  {journal} {Phys. Rev. A}\ }\textbf {\bibinfo {volume}
  {72}},\ \bibinfo {pages} {063619} (\bibinfo {year} {2005})}\BibitemShut
  {NoStop}%
\bibitem [{\citenamefont {Widera}\ \emph {et~al.}(2005)\citenamefont {Widera},
  \citenamefont {Gerbier}, \citenamefont {F\"olling}, \citenamefont {Gericke},
  \citenamefont {Mandel},\ and\ \citenamefont {Bloch}}]{PhysRevLett.95.190405}%
  \BibitemOpen
  \bibfield  {author} {\bibinfo {author} {\bibfnamefont {Artur}\ \bibnamefont
  {Widera}}, \bibinfo {author} {\bibfnamefont {Fabrice}\ \bibnamefont
  {Gerbier}}, \bibinfo {author} {\bibfnamefont {Simon}\ \bibnamefont
  {F\"olling}}, \bibinfo {author} {\bibfnamefont {Tatjana}\ \bibnamefont
  {Gericke}}, \bibinfo {author} {\bibfnamefont {Olaf}\ \bibnamefont {Mandel}},
  \ and\ \bibinfo {author} {\bibfnamefont {Immanuel}\ \bibnamefont {Bloch}},\
  }\bibfield  {title} {\enquote {\bibinfo {title} {Coherent collisional spin
  dynamics in optical lattices},}\ }\href {\doibase
  10.1103/PhysRevLett.95.190405} {\bibfield  {journal} {\bibinfo  {journal}
  {Phys. Rev. Lett.}\ }\textbf {\bibinfo {volume} {95}},\ \bibinfo {pages}
  {190405} (\bibinfo {year} {2005})}\BibitemShut {NoStop}%
\bibitem [{\citenamefont {Widera}\ \emph {et~al.}(2006)\citenamefont {Widera},
  \citenamefont {Gerbier}, \citenamefont {F{\"o}lling}, \citenamefont
  {Gericke}, \citenamefont {Mandel},\ and\ \citenamefont
  {Bloch}}]{1367-2630-8-8-152}%
  \BibitemOpen
  \bibfield  {author} {\bibinfo {author} {\bibfnamefont {Artur}\ \bibnamefont
  {Widera}}, \bibinfo {author} {\bibfnamefont {Fabrice}\ \bibnamefont
  {Gerbier}}, \bibinfo {author} {\bibfnamefont {Simon}\ \bibnamefont
  {F{\"o}lling}}, \bibinfo {author} {\bibfnamefont {Tatjana}\ \bibnamefont
  {Gericke}}, \bibinfo {author} {\bibfnamefont {Olaf}\ \bibnamefont {Mandel}},
  \ and\ \bibinfo {author} {\bibfnamefont {Immanuel}\ \bibnamefont {Bloch}},\
  }\bibfield  {title} {\enquote {\bibinfo {title} {Precision measurement of
  spin-dependent interaction strengths for spin-1 and spin-2 87 {R}b atoms},}\
  }\href {http://stacks.iop.org/1367-2630/8/i=8/a=152} {\bibfield  {journal}
  {\bibinfo  {journal} {New Journal of Physics}\ }\textbf {\bibinfo {volume}
  {8}},\ \bibinfo {pages} {152} (\bibinfo {year} {2006})}\BibitemShut {NoStop}%
\bibitem [{\citenamefont {Gerbier}\ \emph {et~al.}(2006)\citenamefont
  {Gerbier}, \citenamefont {Widera}, \citenamefont {F\"olling}, \citenamefont
  {Mandel},\ and\ \citenamefont {Bloch}}]{PhysRevA.73.041602}%
  \BibitemOpen
  \bibfield  {author} {\bibinfo {author} {\bibfnamefont {Fabrice}\ \bibnamefont
  {Gerbier}}, \bibinfo {author} {\bibfnamefont {Artur}\ \bibnamefont {Widera}},
  \bibinfo {author} {\bibfnamefont {Simon}\ \bibnamefont {F\"olling}}, \bibinfo
  {author} {\bibfnamefont {Olaf}\ \bibnamefont {Mandel}}, \ and\ \bibinfo
  {author} {\bibfnamefont {Immanuel}\ \bibnamefont {Bloch}},\ }\bibfield
  {title} {\enquote {\bibinfo {title} {Resonant control of spin dynamics in
  ultracold quantum gases by microwave dressing},}\ }\href {\doibase
  10.1103/PhysRevA.73.041602} {\bibfield  {journal} {\bibinfo  {journal} {Phys.
  Rev. A}\ }\textbf {\bibinfo {volume} {73}},\ \bibinfo {pages} {041602}
  (\bibinfo {year} {2006})}\BibitemShut {NoStop}%
\bibitem [{\citenamefont {Zhao}\ \emph {et~al.}(2015)\citenamefont {Zhao},
  \citenamefont {Jiang}, \citenamefont {Tang}, \citenamefont {Webb},\ and\
  \citenamefont {Liu}}]{PhysRevLett.114.225302}%
  \BibitemOpen
  \bibfield  {author} {\bibinfo {author} {\bibfnamefont {L.}~\bibnamefont
  {Zhao}}, \bibinfo {author} {\bibfnamefont {J.}~\bibnamefont {Jiang}},
  \bibinfo {author} {\bibfnamefont {T.}~\bibnamefont {Tang}}, \bibinfo {author}
  {\bibfnamefont {M.}~\bibnamefont {Webb}}, \ and\ \bibinfo {author}
  {\bibfnamefont {Y.}~\bibnamefont {Liu}},\ }\bibfield  {title} {\enquote
  {\bibinfo {title} {Antiferromagnetic spinor condensates in a two-dimensional
  optical lattice},}\ }\href {\doibase 10.1103/PhysRevLett.114.225302}
  {\bibfield  {journal} {\bibinfo  {journal} {Phys. Rev. Lett.}\ }\textbf
  {\bibinfo {volume} {114}},\ \bibinfo {pages} {225302} (\bibinfo {year}
  {2015})}\BibitemShut {NoStop}%
\bibitem [{\citenamefont {Koashi}\ and\ \citenamefont
  {Ueda}(2000)}]{PhysRevLett.84.1066}%
  \BibitemOpen
  \bibfield  {author} {\bibinfo {author} {\bibfnamefont {Masato}\ \bibnamefont
  {Koashi}}\ and\ \bibinfo {author} {\bibfnamefont {Masahito}\ \bibnamefont
  {Ueda}},\ }\bibfield  {title} {\enquote {\bibinfo {title} {Exact eigenstates
  and magnetic response of spin-1 and spin-2 {B}ose-{E}instein condensates},}\
  }\href {\doibase 10.1103/PhysRevLett.84.1066} {\bibfield  {journal} {\bibinfo
   {journal} {Phys. Rev. Lett.}\ }\textbf {\bibinfo {volume} {84}},\ \bibinfo
  {pages} {1066--1069} (\bibinfo {year} {2000})}\BibitemShut {NoStop}%
\bibitem [{\citenamefont {Ho}\ and\ \citenamefont
  {Yip}(2000)}]{PhysRevLett.84.4031}%
  \BibitemOpen
  \bibfield  {author} {\bibinfo {author} {\bibfnamefont {Tin-Lun}\ \bibnamefont
  {Ho}}\ and\ \bibinfo {author} {\bibfnamefont {Sung~Kit}\ \bibnamefont
  {Yip}},\ }\bibfield  {title} {\enquote {\bibinfo {title} {Fragmented and
  single condensate ground states of spin-1 {B}ose gas},}\ }\href {\doibase
  10.1103/PhysRevLett.84.4031} {\bibfield  {journal} {\bibinfo  {journal}
  {Phys. Rev. Lett.}\ }\textbf {\bibinfo {volume} {84}},\ \bibinfo {pages}
  {4031--4034} (\bibinfo {year} {2000})}\BibitemShut {NoStop}%
\bibitem [{\citenamefont {Deuchert}\ \emph {et~al.}(2012)\citenamefont
  {Deuchert}, \citenamefont {Sakmann}, \citenamefont {Streltsov}, \citenamefont
  {Alon},\ and\ \citenamefont {Cederbaum}}]{PhysRevA.86.013618}%
  \BibitemOpen
  \bibfield  {author} {\bibinfo {author} {\bibfnamefont {Andreas}\ \bibnamefont
  {Deuchert}}, \bibinfo {author} {\bibfnamefont {Kaspar}\ \bibnamefont
  {Sakmann}}, \bibinfo {author} {\bibfnamefont {Alexej~I.}\ \bibnamefont
  {Streltsov}}, \bibinfo {author} {\bibfnamefont {Ofir~E.}\ \bibnamefont
  {Alon}}, \ and\ \bibinfo {author} {\bibfnamefont {Lorenz~S.}\ \bibnamefont
  {Cederbaum}},\ }\bibfield  {title} {\enquote {\bibinfo {title} {Dynamics and
  symmetries of a repulsively bound atom pair in an infinite optical
  lattice},}\ }\href {\doibase 10.1103/PhysRevA.86.013618} {\bibfield
  {journal} {\bibinfo  {journal} {Phys. Rev. A}\ }\textbf {\bibinfo {volume}
  {86}},\ \bibinfo {pages} {013618} (\bibinfo {year} {2012})}\BibitemShut
  {NoStop}%
\bibitem [{\citenamefont {Winkler}\ \emph {et~al.}(2006)\citenamefont
  {Winkler}, \citenamefont {Thalhammer}, \citenamefont {Lang}, \citenamefont
  {Grimm}, \citenamefont {Hecker~Denschlag}, \citenamefont {Daley},
  \citenamefont {Kantian}, \citenamefont {B{\"u}chler},\ and\ \citenamefont
  {Zoller}}]{Winkler2006RB}%
  \BibitemOpen
  \bibfield  {author} {\bibinfo {author} {\bibfnamefont {K.}~\bibnamefont
  {Winkler}}, \bibinfo {author} {\bibfnamefont {G.}~\bibnamefont {Thalhammer}},
  \bibinfo {author} {\bibfnamefont {F.}~\bibnamefont {Lang}}, \bibinfo {author}
  {\bibfnamefont {R.}~\bibnamefont {Grimm}}, \bibinfo {author} {\bibfnamefont
  {J.}~\bibnamefont {Hecker~Denschlag}}, \bibinfo {author} {\bibfnamefont
  {A.~J.}\ \bibnamefont {Daley}}, \bibinfo {author} {\bibfnamefont
  {A.}~\bibnamefont {Kantian}}, \bibinfo {author} {\bibfnamefont {H.~P.}\
  \bibnamefont {B{\"u}chler}}, \ and\ \bibinfo {author} {\bibfnamefont
  {P.}~\bibnamefont {Zoller}},\ }\bibfield  {title} {\enquote {\bibinfo {title}
  {Repulsively bound atom pairs in an optical lattice},}\ }\href
  {http://dx.doi.org/10.1038/nature04918} {\bibfield  {journal} {\bibinfo
  {journal} {Nature}\ }\textbf {\bibinfo {volume} {441}},\ \bibinfo {pages}
  {853--856} (\bibinfo {year} {2006})}\BibitemShut {NoStop}%
\bibitem [{\citenamefont {Folling}\ \emph {et~al.}(2007)\citenamefont
  {Folling}, \citenamefont {Trotzky}, \citenamefont {Cheinet}, \citenamefont
  {Feld}, \citenamefont {Saers}, \citenamefont {Widera}, \citenamefont
  {Muller},\ and\ \citenamefont {Bloch}}]{Folling:2007aa}%
  \BibitemOpen
  \bibfield  {author} {\bibinfo {author} {\bibfnamefont {S.}~\bibnamefont
  {Folling}}, \bibinfo {author} {\bibfnamefont {S.}~\bibnamefont {Trotzky}},
  \bibinfo {author} {\bibfnamefont {P.}~\bibnamefont {Cheinet}}, \bibinfo
  {author} {\bibfnamefont {M.}~\bibnamefont {Feld}}, \bibinfo {author}
  {\bibfnamefont {R.}~\bibnamefont {Saers}}, \bibinfo {author} {\bibfnamefont
  {A.}~\bibnamefont {Widera}}, \bibinfo {author} {\bibfnamefont
  {T.}~\bibnamefont {Muller}}, \ and\ \bibinfo {author} {\bibfnamefont
  {I.}~\bibnamefont {Bloch}},\ }\bibfield  {title} {\enquote {\bibinfo {title}
  {Direct observation of second-order atom tunnelling},}\ }\href
  {http://dx.doi.org/10.1038/nature06112} {\bibfield  {journal} {\bibinfo
  {journal} {Nature}\ }\textbf {\bibinfo {volume} {448}},\ \bibinfo {pages}
  {1029--1032} (\bibinfo {year} {2007})}\BibitemShut {NoStop}%
\bibitem [{\citenamefont {Konno}(2005)}]{PhysRevE.72.026113}%
  \BibitemOpen
  \bibfield  {author} {\bibinfo {author} {\bibfnamefont {Norio}\ \bibnamefont
  {Konno}},\ }\bibfield  {title} {\enquote {\bibinfo {title} {Limit theorem for
  continuous-time quantum walk on the line},}\ }\href {\doibase
  10.1103/PhysRevE.72.026113} {\bibfield  {journal} {\bibinfo  {journal} {Phys.
  Rev. E}\ }\textbf {\bibinfo {volume} {72}},\ \bibinfo {pages} {026113}
  (\bibinfo {year} {2005})}\BibitemShut {NoStop}%
\bibitem [{\citenamefont {Breit}\ and\ \citenamefont
  {Rabi}(1931)}]{PhysRev.38.2082.2}%
  \BibitemOpen
  \bibfield  {author} {\bibinfo {author} {\bibfnamefont {G.}~\bibnamefont
  {Breit}}\ and\ \bibinfo {author} {\bibfnamefont {I.~I.}\ \bibnamefont
  {Rabi}},\ }\bibfield  {title} {\enquote {\bibinfo {title} {Measurement of
  nuclear spin},}\ }\href {\doibase 10.1103/PhysRev.38.2082.2} {\bibfield
  {journal} {\bibinfo  {journal} {Phys. Rev.}\ }\textbf {\bibinfo {volume}
  {38}},\ \bibinfo {pages} {2082--2083} (\bibinfo {year} {1931})}\BibitemShut
  {NoStop}%
\end{thebibliography}%

\end{document}